# Electrochemistry-Enhanced Dynamic Paths Sampling Unveiling Nuclear Quantum Effects in Electrocatalysis

Li Fu[1], Yifan Li[2], Menglin Sun[1], Xiaolong Yang[1], Bin Jin[1], Shenzhen Xu[1]*

[1]Beijing Key Laboratory of Theory and Technology for Advanced Battery Materials, School of Materials Science and Engineering, Peking University, Beijing 100871, People's Republic of China

[2]Department of Chemistry, Princeton University, Princeton, NJ 08544, USA

*Corresponding author: xushenzhen@pku.edu.cn

## Abstract

Proton-coupled electron transfers (PCET) are elementary steps in electrocatalysis. However, accurate calculations of PCET rates remain challenging, especially considering nuclear quantum effects (NQEs) under a constant potential condition. Statistical sampling of reaction paths is an ideal approach for rate calculations, however, is always limited by the rare-event issue. Here we develop an electrochemistry-driven quantum dynamics approach enabling realistic enhanced paths sampling under constant potentials without *a priori* defined reaction coordinates. We apply the method in modeling the Volmer step of the hydrogen evolution reaction, and demonstrate that the NQEs exhibit more than one order of magnitude impact on the computed rate constant, indicating an essential role of NQEs in electrochemistry.

Proton-coupled electron transfer (PCET) reactions are critical elementary steps in many electrochemical processes and play significant roles in electrocatalytic systems [1–4]. Accurate computation of PCET rate constants is essential for mechanistic understanding and optimizing our interested electrochemical processes. Previous computational studies [5–9] typically constructed potential energy landscapes or free energy profiles of PCET steps, and roughly estimated the rate constants based on the Transition State Theory (TST) [10,11], without dynamic paths sampling and neglecting the dynamic recrossing effects around transition states [12]. To address this limitation, the Bennett-Chandler (BC) approach [13,14] incorporates dynamic corrections by evaluating the flux-side time-correlation function [15,16], offering a more rigorous treatment of dynamic contributions [12,17–19]. However, the required free energy calculations still rely on *a priori* selection of reaction coordinates (RCs), introducing artificial assumptions that may affect the reliability of calculated rate constants. Thus, a computational method without predefined RCs becomes desirable. One straightforward strategy is running multiple trajectories and statistically analyzing the time required for a system's transition from its initial state (IS) to its final state (FS), which is equivalent to the computation of the side-side time-correlation function by dynamic paths sampling [12,20–22]. However, this method requires sampling a large number of trajectories over long time scales to accurately capture the system's dynamics, especially for high-barrier reactions. Furthermore, it is necessary to choose a specific force field for the simulations. One possible choice is *ab-initio* molecular dynamics (AIMD) [23,24], but it requires substantial computational costs. An alternative option is using machine learning force fields (MLFF) [25,26], which greatly reduces the burden of energy computation. However, the cost would remain significant even with MLFF, particularly for modeling rare-event dynamics.

Regarding the challenges, the issues could be resolved in electrochemical systems. Electrocatalytic surface reactions in experiments inherently occur in open systems at constant electrode potential, corresponding to a grand canonical ensemble where electrons exchange with a reservoir [27–30]. It is well known that external electrode potential can accelerate electrochemical reactions in realistic experiments. Thus, the electrochemical driving force might perform as a knob to achieve realistic enhanced sampling for dynamic paths in the side-side time-correlation function calculations [12,20–22], unlike the artificial bias commonly used in enhanced sampling simulations [31–33]. More specifically, we could compute electro-reduction/oxidation PCET rate constants at potentials with strong reducing/oxidizing power, leading to relatively low barriers and only requesting affordable cost. We can utilize the Tafel relationship [34–37] in electrochemistry to extrapolate to our interested bias potential conditions that may exhibit high kinetic barriers, by which we could solve the rare-event sampling issue.

Earlier theoretical studies [5–9] in electrocatalysis often ignored nuclear quantum effects (NQEs) in PCET simulations. However, NQEs associated with the transferring protons could be important. Much evidence suggests that NQEs remain influential even at room temperature, particularly in processes involving proton transfers in liquid water [38], biomolecules [39], or even crystalline oxide materials [40]. Additionally, a recent computational study by Sun et al [29]. reported that NQEs lead to non-negligible reduction in activation energies of PCET steps in the hydrogen evolution reaction (HER). These findings call for a revisit of the role of NQEs in electrochemical processes, emphasizing the need of incorporating NQEs to achieve a more accurate description of PCET dynamics. However, existing molecular simulation methods only separately consider implementing a constant potential condition in classical molecular dynamics (MD), or incorporating NQEs in the canonical ensemble simulations. In classical MD, a constant potential condition is typically achieved by introducing a potentiostat [27,30], which controls the electrochemical potential to fluctuate around a constant value. When both temperature and electrochemical potential are controlled, the approach is referred to as thermostatted-potentiostatted classical MD (TP-Classical MD) [27]. The extended Nosé–Hoover Hamiltonian of the TP-Classical MD algorithm is described in the **Supplementary Information (SI) Section 1**. The GC ensemble partition function can be expressed as

$$\Xi(\beta, \mu_{\mathrm{e}}) = \sum_{N_{\mathrm{e}}} \Lambda \int \mathrm{d}\boldsymbol{r} \exp\left[-\beta\left(U(\{\boldsymbol{r}_i\}, N_{\mathrm{e}}) - \mu_{\mathrm{e}} N_{\mathrm{e}}\right)\right] = \sum_{N_{\mathrm{e}}} \exp\left(\beta \mu_{\mathrm{e}} N_{\mathrm{e}}\right) Q_{can}(\beta, N_{\mathrm{e}}) \tag{1}$$

where $\beta = \frac{1}{k_{\mathrm{B}}T}$ is the inverse temperature, and $\Lambda$ arises from integrating out the momenta. The potential energy $U(\{\boldsymbol{r}_i\}, N_{\mathrm{e}})$ depends on particle positions $\{\boldsymbol{r}_i\}$ ($\boldsymbol{R} = \{\boldsymbol{r}_i\}$ for simplicity) and total electron number $N_{\mathrm{e}}$.The electrochemical potential $\mu_{\mathrm{e}}$ (a macroscopic quantity) characterizes the electronic reservoir in equilibrium with the modeled system, and $Q_{can}(\beta, N_{\mathrm{e}})$ refers to the canonical ensemble partition function. Under exact GC ensemble conditions, the instantaneous work function $W_{\mathrm{e}}(\{\boldsymbol{r}_i\}, N_{\mathrm{e}})$ of each sampled microstate should exhibit thermal fluctuations near $-\mu_{\mathrm{e}}$, rather than being fixed at a specific value, and satisfy:

$$\langle \frac{\partial U(\{\boldsymbol{r}_i\}, N_e)}{\partial N_e} \rangle_{GC} = \langle -W_e(\{\boldsymbol{r}_i\}, N_e) \rangle_{GC} = \mu_e \tag{2}$$

where $\langle \dots \rangle_{GC}$ denotes the statistical GC ensemble average. Since the electronic chemical potential $\mu_e$ is referenced to vacuum level, it is straightforward to relate the applied potential to a value referenced to the Standard Hydrogen Electrode (~ -4.4 eV vs. vacuum level). **Figure 1(a)** shows a schematic plot of a classical trajectory, illustrating the dynamics of classical particles under electrochemical conditions, coupled to both a thermostat and a potentiostat.

To quantitatively account for NQEs, we employ the path integral (PI) formulation [41–43]. In this framework, the entire system with $N_{total}$ particles is represented by a ring-polymer morphology consisting of $P$ discrete beads connected by harmonic interactions (each bead containing $3N_{total}$ degrees of freedom). The positions of the particles in the $k$-th bead can be denoted as $\boldsymbol{R}^{(k)}$ for simplicity, where $\boldsymbol{R}^{(k)} = \{\boldsymbol{r}_i^{(k)}\}$, with $i$ representing the $i$-th atom and $k$ indicating the $k$-th bead. Thus, $\{\boldsymbol{R}^{(j)}\}$ represents the set of coordinates of all $P$ beads. Two widely used methods for quantum dynamics simulations are the centroid molecular dynamics (CMD) [44] and the ring-polymer molecular dynamics (RPMD) [45]. While both utilize the imaginary-time PI formalism, they differ in implementation: RPMD employs Newtonian dynamics in the extended phase space of the ring polymer, whereas CMD evolves the centroid on a potential of mean force contributed by the internal modes of PI beads. Both methods have limitations: RPMD suffers from spurious resonances and non-ergodicity issues; CMD introduces a curvature problem causing unphysical red shifts and spectral broadening [21,22]. To resolve these issues, Rossi et al. [21] proposed "thermostatted RPMD" (T-RPMD), a hybrid approach that combines RPMD and CMD. T-RPMD retains RPMD bead masses for short-time accuracy and introduces a Langevin thermostat to the internal modes, excluding the centroid mode, thereby eliminating spurious resonances, achieving a more reasonable statistical distribution of microstates, and preserving centroid dynamics. Additionally, T-RPMD mitigates the non-ergodicity issue of RPMD and removes the strict time step constraints of CMD. Hele and Suleimanov [22] further evaluated the performance of T-RPMD in thermal reaction rate calculations, demonstrating its numerical robustness and practical utility.

Inspired by T-RPMD, we extend this approach to electrochemical systems by coupling the potentiostat to the internal modes, resulting in the thermostatted-potentiostatted ring-polymer molecular dynamics (TP-RPMD), as shown in **Figure 1(b)**. Note that the beads in the diagram are visualization tools, while the actual coupling occurs with the modes obtained after the normal mode transformation [43,46,47]. A detailed derivation and dynamical equations are provided in the **SI Section 3**. Since all beads in the ring-polymer share the same electron number $N_\mathrm{e}$ in the PI formalism (further discussion shown in **SI Section 2**), the entire quantized system is coupled to a single potentiostat.

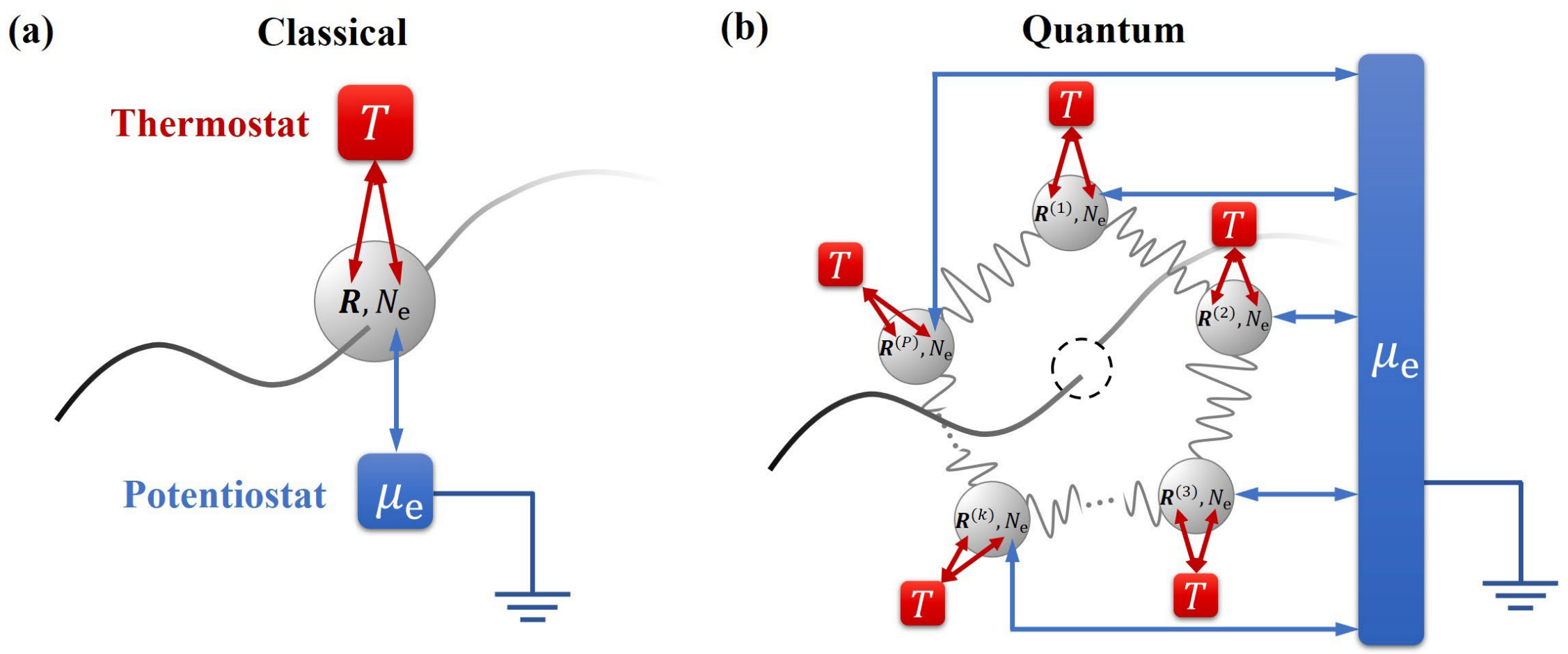

**Figure 1**. (a) Scheme of TP-Classical MD. (b) Scheme of TP-RPMD incorporating NQEs under constant potential conditions. The PI algorithm is illustrated by multiple beads connected by harmonic springs. Black curves: MD Trajectories. Red rectangles: Thermostats coupled to $\boldsymbol{R}$ and $N_\mathrm{e}$. Blue rectangles: Potentiostats coupled to $N_\mathrm{e}$.

With the molecular simulation strategy TP-RPMD, the reaction rate constants can be calculated by the side-side time correlation function, without relying on a specific selection of RCs. For a model system, we define a reactant region $A$ (or IS equivalently) and a product region $B$ (or FS) in the phase space. The side-side time correlation function is defined as [12]:

$$C_{AB}(t) = \frac{\langle h_A(x_0)h_B(x_t)\rangle}{\langle h_A(x_0)\rangle} \approx \langle h_B\rangle \left(1 - e^{-\frac{t}{\tau_R}}\right) \tag{3}$$

The ensemble averages can be written as an integration over the phase-space coordinates at $t = 0$, weighted by the equilibrium ensemble distribution $\mathcal{N}(x_0)$.

$$\langle h_A(x_0)h_B(x_t)\rangle = \frac{\int dx_0 \mathcal{N}(x_0) h_A(x_0) h_B(x_t)}{\int dx_0 \mathcal{N}(x_0)} \tag{4}$$

$$\langle h_A(x_0)\rangle = \frac{\int dx_0 \mathcal{N}(x_0) h_A(x_0)}{\int dx_0 \mathcal{N}(x_0)}$$

where $x_t$ denotes the system's phase-space coordinates $\{\boldsymbol{r}_i, \boldsymbol{p}_i\}$ at time $t$, $h_A$ and $h_B$ are indicator functions ($h_A = 1$ if the system is in the $A$ region, and $h_A = 0$ otherwise; a similar definition holds for $h_B$).

$$h_{A,B}(x) = \begin{cases} 1, & \text{if } x \text{ in } A \text{ or } B \\ 0, & \text{otherwise.} \end{cases} \tag{5}$$

Note that $h_{A,B}$ are not RCs; rather, they serve to indicate whether the system is in IS or FS. $\tau_R$ denotes the reaction time constant, which characterizes the transition timescale between regions. Specifically, $k_{A\rightarrow B}$ and $k_{B\rightarrow A}$ are the forward ($A\rightarrow B$) and the reverse ($B\rightarrow A$) rate constants. The reaction time constant is given by [12]:

$$\tau_R = (k_{A\rightarrow B} + k_{B\rightarrow A})^{-1} \tag{6}$$

$C_{AB}(t)$ measures the conditional probability that a trajectory starting in $A$ at $t = 0$ arrives in $B$ at time $t$. We calculate $\tau_R$ through linear fitting of $\ln\big(1 - C_{AB}(t)\big)$ with respect to time $t$. In our work, we approximate $\langle h_B\rangle$ as unity (further explained by **Eq. 8**). To select an appropriate time range $[0, t^*]$ for the fitting procedure, we maximize the Pearson correlation coefficient ($R^2$) over multiple $t^*$ values, ensuring a robust determination of the reaction rate constant [48].

As discussed above, we utilize the Tafel relationship [34–37] in electrochemistry to effectively achieve an enhanced sampling. Reaction rate constants are initially computed at highly negative applied potentials for the electro-reduction Volmer step. We then do linear extrapolations to predict the PCET reaction rates at our targeted potentials, which can be expressed as:

$$\log_{10}\big(k_{\text{PCET}}(\mu_{\text{e}})\big) = A\mu_{\text{e}} + B \tag{7}$$

The constants $A$ and $B$ are the coefficients of the linear relationship.

The Volmer reaction ($\text{H}^+_{\text{sol}} + \text{e}^- +^* \rightarrow \text{H}^*$), an elementary step in HER, is a fundamental PCET process occurring at electrode surfaces. The Volmer step involves a solvated proton ($\text{H}^+_{\text{sol}}$), an electron ($\text{e}^-$), and a surface site (*), making it an ideal system to demonstrate our electrochemistry-enhanced classical/quantum dynamics method. We use the Pt (111) surface, a well-known catalytic facet for HER, to construct our atomic model for MD simulations [8,49–53]. The structural details are shown in **Figure 2(a)**.

(a)

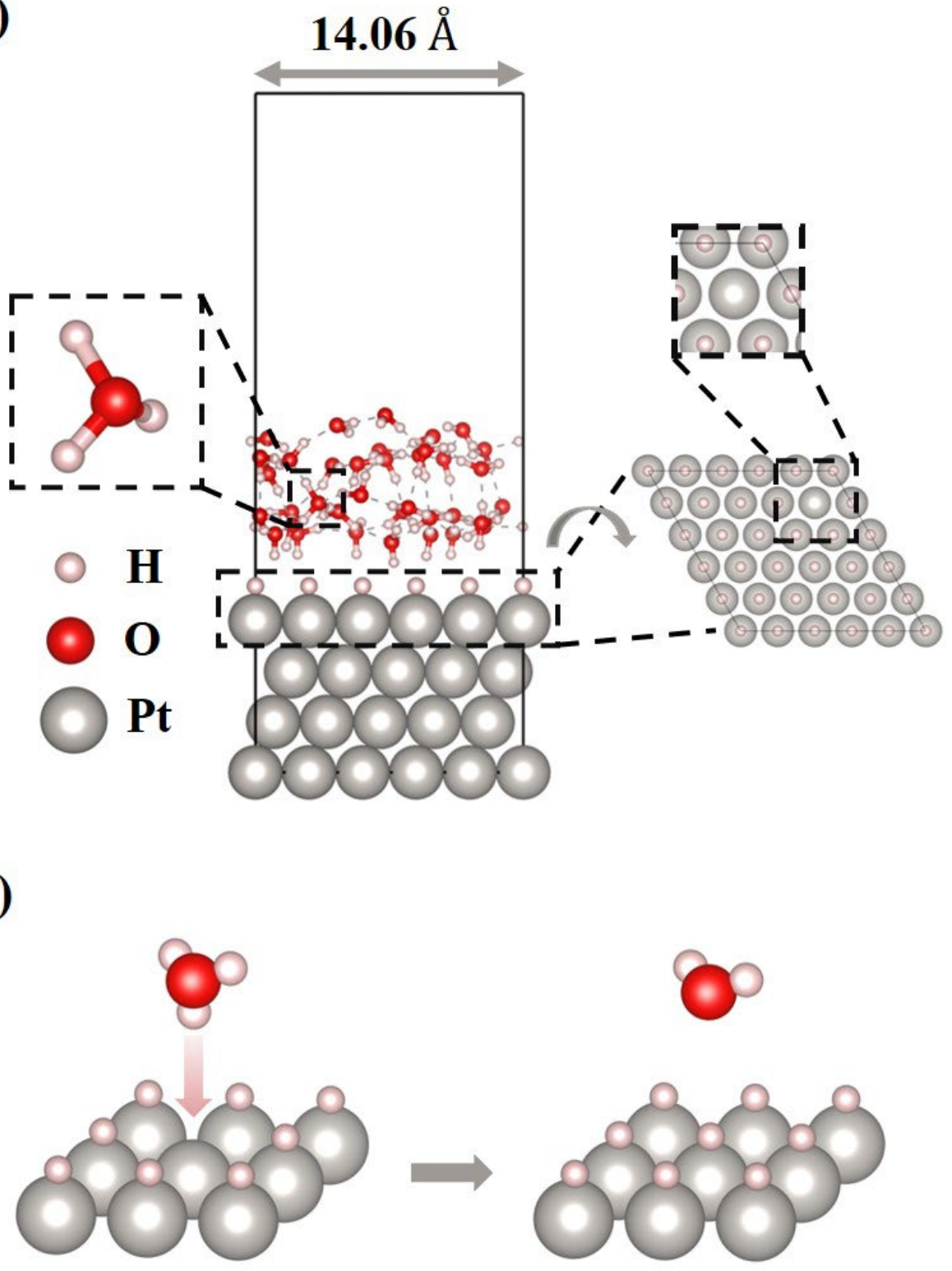

**Figure 2**. (a) Side view of the (5×5) Pt (111) surface slab, consisting of four atomic layers. The bottom two layers of Pt atoms are fixed during MD simulations. The interfacial model includes a water bilayer (36 explicit water molecules) placed over the surface, with a solvated proton (i.e., $H_3O^+$) in the water, whose zoom-in image is shown on the left. One monolayer (1 ML) of hydrogen atoms adsorb at the Pt surface, with one site unoccupied (the top view shown on the right), thus enabling a proton to transfer to a surface site in the FS. (b) Scheme of the Volmer step's pathway.

Given the atomic interfacial model for the Volmer reaction, we employ MLFF to accelerate our molecular simulations (details of MLFF training are shown in the **SI Section 5**). Both the TP-Classical MD and TP-RPMD simulations are performed using the LAMMPS [54] package. The time step for integrating equations of motion is set as 0.5 fs. We perform MD simulations under a GC ensemble condition, with temperature at 300 K and applied potentials controlled at different $\mu_e$ values. We employ 16 beads in the quantum dynamics simulations [29,55,56]. We first perform normal mode TP-PIMD simulations for configurations sampling in the IS region, then conduct TP-RPMD based on the above sampled configurations as the MD starting points to investigate the dynamic behavior $C_{AB}(t)$ of our modeled Volmer reaction step.

To define the FS region of the Volmer reaction, we track positions of protons in the water solvation layer, and check if a certain proton transfers to the Pt surface becoming an adsorbed hydrogen. Specifically, if a hydrogen atom (originally belonging to water solvation) is at a position with its nearest neighboring Pt atom of a Pt-H distance less than 1.7 Å (the fully relaxed Pt-H bond is ~ 1.5 – 1.6 Å [9,57,58]) and its nearest neighboring O atom of an O-H distance larger than 1.6 Å (the O-H bond length in a solvated proton $H_3O^+$ is ~ 1.0 – 1.1 Å [9,59]), this hydrogen is considered to be transferred from water onto the Pt surface, indicating that our modeled system evolves into a FS region. In addition, since we study a PCET step, the FS exhibits a higher $N_\mathrm{e}$ than the IS under both classical and quantum conditions, as shown in **Figure 3(a, b)**. Our FS identification criterion has the advantage that it does not require choosing a specific proton and track its position, as commonly adopted in earlier computational electrochemistry studies [9,60–62] based on predefined RCs. In contrast, our approach treats all protons in the water layer equivalently.

In addition to defining the range of the FS region, we still need an additional criterion to determine if the system truly resides in the FS region. We check the system's state along a dynamic trajectory and confirm the system being in the FS region if it spends more than half of the time within 100 successive steps in the FS region. We use centroids of beads as atomic positions in quantum cases to identify the system's reaction state. Examples illustrating this criterion for both classical and quantum conditions are shown in **Figure S2**. This criterion can be rationalized by the results shown in **Figure 3(a, b)**, where the moment of the system's transition to the FS region (marked by the vertical red dashed line), determined by the above introduced strategy, well matches with the step-like jump of $N_\mathrm{e}$ (another physically meaningful indicator for the system's state identification). Once a trajectory is confirmed as falling into the FS region, we terminate the simulation under an assumption of no reverse reactions from the FS region (H*) to the IS region ($H^+_{sol}$ + $e^-$ + *), i.e., $k_{B\to A} = 0$, and the reaction rate constant $k_{A\to B} = \tau_R{}^{-1}$, which also means we assume $\langle h_B \rangle = 1$ at equilibrium in **Eq. 3**. To simplify the notation in subsequent expressions, we use $k_\mathrm{PCET}$ to represent $k_{A\to B}$, then **Eq. 3** becomes:

$$C_{AB}(t) = 1 - e^{-k_\mathrm{PCET}t} \tag{8}$$

We calculate the reaction rate constants at each applied $\mu_\mathrm{e}$ using the methods described. For classical situations, 200 trajectories are used for sampling, whereas for the quantum cases, 100 trajectories are used, considering the computational cost. The $k_{\text{PCET}}$ fitting results based on dynamic sampling of the side-side time correlation function are presented in **Figure 3(c, d)**. The time range $t^*$ chosen for this linear fitting yields the maximum $R^2$ value (refer to **SI Section 7**).

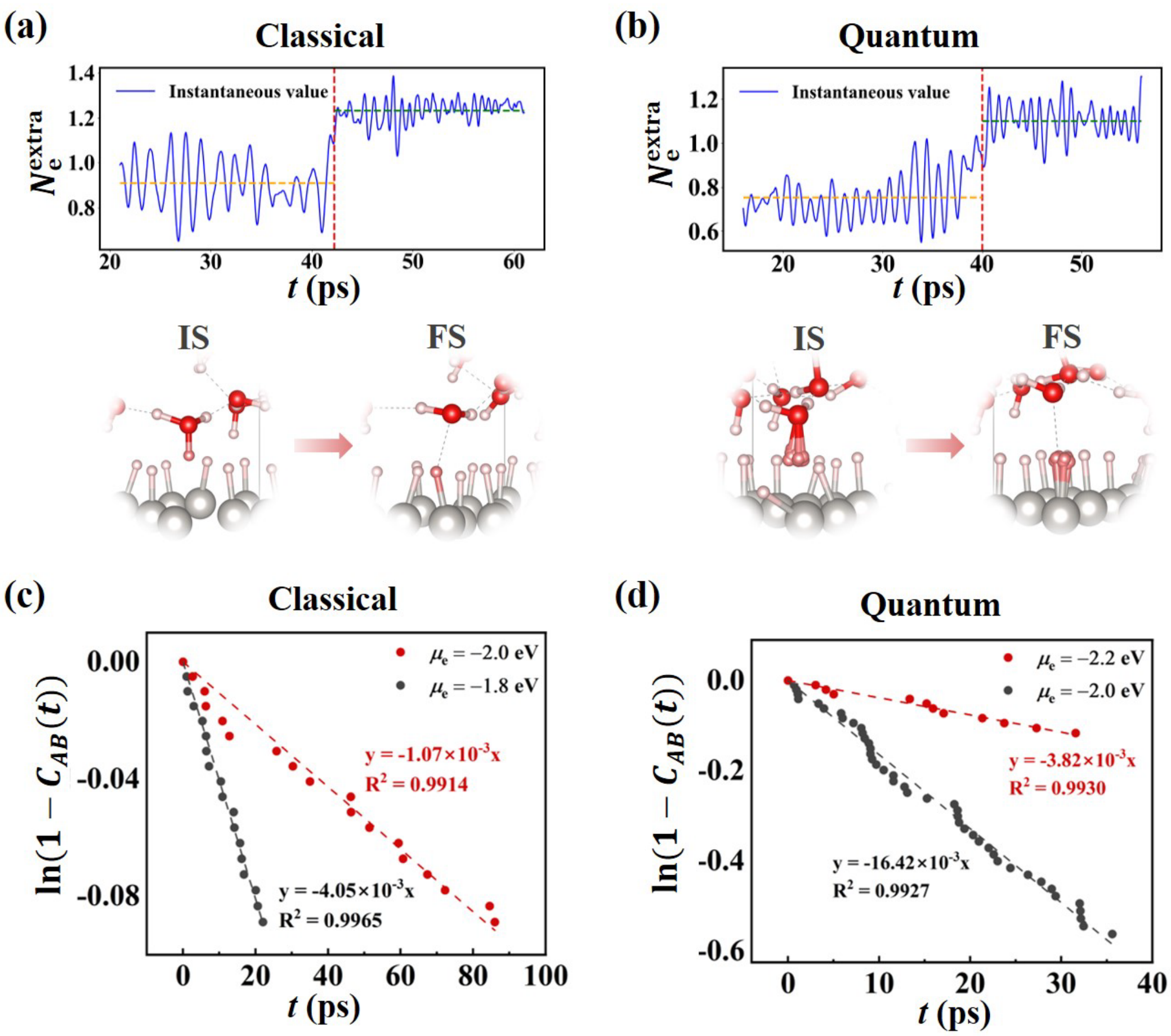

**Figure 3**. (a) Total electron number $N_{\text{e}}$ variation with respect to MD time steps, associated with schematic structural plots of IS and FS under classical situations. The extra electron number ($N_{\text{e}}^{\text{extra}}$) added or removed from a neutral model represents $N_{\text{e}}$ fluctuation for clarity. Instantaneous values of $N_{\text{e}}^{\text{extra}}$ are shown by the blue curve with a vertical red dashed line marking the system's transition moment into the Volmer reaction's FS. The transferred hydrogen atom is highlighted by dark-colored spheres. (b) The same convention as (a), but under quantum situations instead, here we only show the expanded PI beads of the transferred hydrogen atom for simplicity and clarity. (c) Time evolution of $\ln(1 - C_{AB}(t))$ under classical situations at $\mu_{\text{e}}$ = -1.8 eV, the corresponding $k_{\text{PCET}}$ = 4.05 ns$^{-1}$ and at $\mu_{\text{e}}$ = -2.0 eV, the corresponding $k_{\text{PCET}}$ = 1.07 ns$^{-1}$. (d) The same convention as (c), but under quantum situations instead, $k_{\text{PCET}}$ = 16.42 ns$^{-1}$ at $\mu_{\text{e}}$ = -2.0 eV, and $k_{\text{PCET}}$ = 3.82 ns$^{-1}$ at $\mu_{\text{e}}$ = -2.2 eV.

The Volmer PCET rate constants at different $\mu_e$ conditions are presented in **Figure 4(a)**. TP-Classical MD under highly reducing conditions (from -1.6 eV to -2.2 eV vs. vacuum) shows a linear correlation between $\log_{10}(k_{PCET}(\mu_e))$ and $\mu_e$. As shown in **Figure 4(a)**, a computationally demanding test point at $\mu_e$ = -2.5 eV (red star) falls exactly on the extrapolated blue dashed line, justifying the linear fitting strategy based on the Tafel relationship. In addition, we run TP-RPMD at three different potentials (from -2.0 eV to -2.4 eV vs. vacuum) to investigate the effects of NQEs on the Volmer PCET reaction. The good linear feature of $\log_{10}(k_{PCET}(\mu_e))$ with respect to $\mu_e$ in the quantum situation further validates our electrochemistry-driven realistic enhanced sampling strategy.

When considering NQEs in quantum dynamics, the spreading range of expanded beads (relative to the centroid) of the transferring proton could provide qualitative insights into the proton tunneling effect, which is shown in the upper panel of **Figure 4(b)**. The distance difference between the Pt-H and O-H pairs (where H is the transferring proton), calculated using the beads' centroids as atomic positions, serves as the order parameter, as shown in the lower panel of **Figure 4(b)**. We note that this configurational order parameter is not used as a RC, but simply as an indicator of the system's state along a PCET path. By comparing the upper and lower panels of **Figure 4(b)**, we observe that the transferring proton's beads tend to spread with a larger spatial range right at the system's transition moment to the Volmer step's FS, directly indicating a proton tunneling behavior.

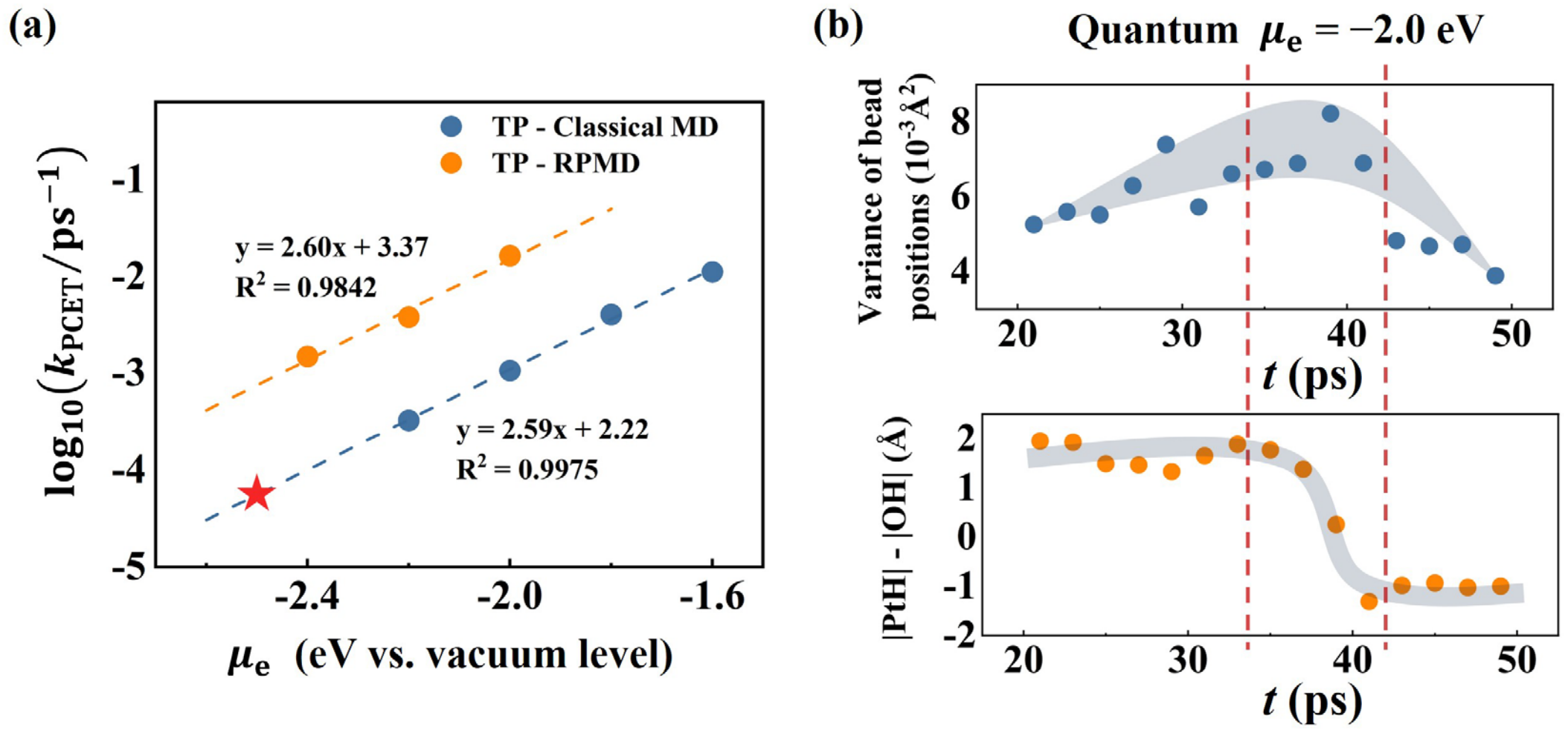

**Figure 4**. (a) Linear dependence of $\log_{10}(k_{\mathrm{PCET}}(\mu_{\mathrm{e}}))$ on electrochemical potentials denoted by $\mu_{\mathrm{e}}$ referenced to vacuum level, where $k_{\mathrm{PCET}}$ is in the unit of $ps^{-1}$. Blue data points: TP-Classical MD (used for linear fitting); orange data points: TP-RPMD; Red star: TP-Classical MD at $\mu_{\mathrm{e}}$ = -2.5 eV, validating the linear extrapolation scheme. (b) Upper panel: time evolution of the spreading range of the transferring proton's beads along the z-direction normal to the Pt surface. The variance of bead positions is obtained by computing the mean square error of the beads' z-coordinates relative to the centroid. Lower panel: difference between the Pt-H and O-H pairs' distances, where H is the transferring proton, Pt and O are the neighboring atoms. This quantity performs as a configurational order parameter monitoring the reaction progress. Each data point in both panels represents an average over 5000 MD steps (2.5 ps). Grey-colored regions are used as guidelines illustrating the general trend. The red dashed vertical lines indicate the time period of the system's transition to the FS region.

Our group's earlier work [29] computed the Volmer step's free energy profile at the potential of $\mu_{\mathrm{e}}$ = -3.5 eV and obtained the thermodynamic activation energies for both classical and quantum situations. By applying the TST equation, we can roughly estimate the impact of NQEs on the computed rate constants, which differs by 1-2 orders of magnitude. The extrapolated rate constants in this work are $k_{\mathrm{PCET}}$(classical) = $1.5\times10^{-4}$ $ns^{-1}$ and $k_{\mathrm{PCET}}$ (quantum) = $1.9\times10^{-3}$ $ns^{-1}$ at $\mu_{\mathrm{e}}$ = -3.5 eV from the results shown in **Figure 4(a)**. Consistent with the thermodynamic analysis from our earlier study, **the protons' NQEs indeed exhibit about ≥1 order of magnitude impact on the computed PCET rate constants**, which is obviously non-negligible even at room temperature. Our theoretical work thus highlights the physical importance of NQEs in electrochemistry, which is worth being explored in broader application scenarios of electrocatalytic oxygen reduction/evolution reactions, $CO_2$ reduction reactions, and nitrogen reduction reactions in the future.

Another highlighted point of our work is the strategy to achieve realistic enhanced sampling by utilizing the Tafel relationship in electrochemistry, paving the way to address the challenges of rare-event dynamics without any artificial bias or *a priori* RCs. Generally, enhanced sampling methods can be classified into two major categories: collective variable (CV)-based approaches (e.g. metadynamics [63], umbrella sampling [64,65]), and methods that modify the macroscopic thermodynamic conditions of the system (e.g., parallel tempering [33,66–68] which involves multiple ensembles at various temperatures). Our enhanced sampling approach falls within the second category, where we tune the macroscopic condition – the electrochemical potential $\mu_{\mathrm{e}}$ – of the modeled system to achieve efficient dynamic paths sampling at electrochemical conditions with relatively low barriers, and then extrapolate to the potential conditions with sluggish kinetics to obtain the rate constants based on the Tafel relationship. The major advantage is that $\mu_e$ in the simulations exactly corresponds to the macroscopic condition – applied potentials – in realistic experiments. This strategy is analogous to the extrapolated calculation of diffusion coefficients, where diffusivities are computed at high temperatures and extrapolated to lower temperatures via the Arrhenius-type relationship. However, temperature elevation could induce significant structural changes in materials, making extrapolation of physical quantities invalid in certain cases (e.g. an earlier work on Li diffusion [69]). Our electrochemistry-driven realistic enhanced sampling strategy could mitigate this issue, as the variation of $\mu_e$ within our interested potential range is less likely to impose large impact on electrode material's structure, while maintaining an effective solution for the rare-event sampling, thus providing a robust method for studying electrochemical dynamics.

In summary, we develop the TP-RPMD approach to calculate PCET reaction rate constants without any *a priori* defined RCs, accounting for NQEs under a grand canonical constant potential condition. By leveraging MLFF-accelerated MD simulations under the electrochemical driving force and the Tafel relationship, our approach achieves realistic enhanced sampling for rare-event dynamic paths. Notably, by performing both TP-classical MD and TP-RPMD simulations in modeling HER Volmer step on Pt (111) surface, we find that NQEs exhibit a remarkable impact on computed dynamic rate constants by more than an order of magnitude increase compared to classical situations, demonstrating the necessity of explicit treatment of NQEs in broader application scenarios of electrochemical systems involving PCET steps.

The authors acknowledge funding support from the National Natural Science Foundation of China (grant no. 92470114, no. 52273223), Ministry of Science and Technology of the People's Republic of China (grant no. 2021YFB3800303), and DP Technology Corporation (grant no. 2021110016001141). The computing resource of this work was provided by the Bohrium Cloud Platform, which was supported by DP Technology.

# Supplemental Information

# Electrochemistry-Enhanced Dynamic Paths Sampling Unveiling Nuclear Quantum Effects in Electrocatalysis

Li Fu[1], Yifan Li[2], Menglin Sun[1], Xiaolong Yang[1], Bin Jin[1], Shenzhen Xu[1]*

[1]Beijing Key Laboratory of Theory and Technology for Advanced Battery Materials, School of Materials Science and Engineering, Peking University, Beijing 100871, People's Republic of China

[2]Department of Chemistry, Princeton University, Princeton, NJ 08544, USA

## S1. Derivation of the classical dynamic equations under a grand canonical constant potential condition

The extended Nosé–Hoover Hamiltonian of the thermostatted-potentiostatted classical molecular dynamics (TP-Classical MD) method under a constant potential grand canonical (GC) ensemble takes the form:

$$H^{\text{ext}} = \sum_{i=1}^{N} \frac{\boldsymbol{p}_i^2}{2m_i s^2} + U(\{\boldsymbol{r}_i\}, N_\text{e}) + \frac{p_s^2}{2Q} + g k_\text{B} T \ln s + \frac{p_{N_\text{e}}^2}{2m_{N_\text{e}} s^2} - \mu_\text{e} N_\text{e} \tag{S1}$$

Here, $\boldsymbol{r}_1, \dots, \boldsymbol{r}_N$ correspond to particles' coordinates and $\boldsymbol{p}_1, \dots, \boldsymbol{p}_N$ represent the momenta, where the subscripts refer to particles' index. We introduce $s$ as the degree of freedom of a thermostat, and it scales particles' velocities accordingly, with its conjugate momentum denoted as $p_s$ and an associated mass $Q$. $T$ is the temperature of the system. The parameter $g$ is determined by a condition that yields a GC distribution in physical phase space. Under this constant potential framework, the system's total electron number $N_\text{e}$ becomes a dynamic variable, with its conjugate momentum $p_{N_\text{e}}$ and an associated mass $m_{N_\text{e}}$.

The equations of motion generated by $H^{\text{ext}}$ are

$$\begin{cases} \dot{\boldsymbol{r}}_i = \dfrac{\partial H^{\text{ext}}}{\partial \boldsymbol{p}_i} = \dfrac{\boldsymbol{p}_i}{m_i s^2} \\ \dot{\boldsymbol{p}}_i = -\dfrac{\partial H^{\text{ext}}}{\partial \boldsymbol{r}_i} = -\dfrac{\partial U}{\partial \boldsymbol{r}_i} = \boldsymbol{F}_i \\ \dot{s} = \dfrac{\partial H^{\text{ext}}}{\partial p_s} = \dfrac{p_s}{Q} \\ \dot{p}_s = -\dfrac{\partial H^{\text{ext}}}{\partial s} = \displaystyle\sum_{i=1}^{N} \dfrac{\boldsymbol{p}_i^2}{m_i s^3} - \dfrac{g k_\text{B} T}{s} + \dfrac{p_{N_\text{e}}^2}{m_{N_\text{e}} s^3} \\ \dot{N}_\text{e} = \dfrac{\partial H^{\text{ext}}}{\partial p_{N_\text{e}}} = \dfrac{p_{N_\text{e}}}{m_{N_\text{e}} s^2} \\ \dot{p}_{N_\text{e}} = -\dfrac{\partial H^{\text{ext}}}{\partial N_\text{e}} = -\dfrac{\partial U}{\partial N_\text{e}} + \mu_\text{e} = F_{N_\text{e}} \end{cases} \tag{S2}$$

A non-canonical transformation of variables is introduced,

$$\boldsymbol{p}'_i = \frac{\boldsymbol{p}_i}{s}, dt' = \frac{dt}{s}, \frac{1}{s}\frac{\text{d}s'}{\text{d}t'} = \frac{\text{d}\eta}{\text{d}t'}, p_s = p_\eta, p'_{N_\text{e}} = \frac{p_{N_\text{e}}}{s} \tag{S3}$$

which leads to a new set of equations of motion in the form of primed coordinates. Here, $\eta$ and $p_\eta$ represent the generalized coordinate and generalized momentum associated with the Nosé–Hoover thermostat's degrees of freedom. In our work, we redefine the physical quantities by removing the primes, ultimately obtaining the final form presented here:

$$\begin{cases} \dot{\boldsymbol{r}}_i = \dfrac{\boldsymbol{p}_i}{m_i} \\ \dot{\boldsymbol{p}}_i = \boldsymbol{F}_i - \dfrac{p_\eta}{Q}\boldsymbol{p}_i \\ \dot{\eta} = \dfrac{p_\eta}{Q} \\ \dot{p}_\eta = \displaystyle\sum_{i=1}^{N} \frac{\boldsymbol{p}_i^2}{m_i} - g k_{\mathrm{B}} T + \frac{p_{N_{\mathrm{e}}}^2}{m_{N_{\mathrm{e}}}} \\ \dot{N}_{\mathrm{e}} = \dfrac{p_{N_{\mathrm{e}}}}{m_{N_{\mathrm{e}}}} \\ \dot{p}_{N_{\mathrm{e}}} = -\dfrac{\partial U}{\partial N_{\mathrm{e}}} + \mu_{\mathrm{e}} - \dfrac{p_\eta}{Q} p_{N_{\mathrm{e}}} \end{cases} \tag{S4}$$

We then compute the physical phase space distribution generated by the above equations of motion (**Eq. S4**). We first need to compute the phase-space compressibility as follows:

$$\kappa = \sum_{i=1}^{N} \left(\nabla_{r_i}\dot{\boldsymbol{r}}_i + \nabla_{\boldsymbol{p}_i}\dot{\boldsymbol{p}}_i\right) + \frac{\partial \dot{\eta}}{\partial \eta} + \frac{\partial \dot{p}_\eta}{\partial p_\eta} + \frac{\partial \dot{N}_{\mathrm{e}}}{\partial N_{\mathrm{e}}} + \frac{\partial \dot{p}_{N_{\mathrm{e}}}}{\partial p_{N_{\mathrm{e}}}} = -\sum_{i=1}^{N} \frac{3p_\eta}{Q} - \frac{p_\eta}{Q} = -(3N+1)\dot{\eta} \tag{S5}$$

Next, we recognize that the phase-space compressibility $\kappa$ is related to the time derivative of $w$

$$\kappa = \frac{\mathrm{d}}{\mathrm{d}t} w \tag{S6}$$

From this relationship, we can set the phase-space metric as:

$$\exp(-w) = \exp[(3N+1)\eta] \tag{S7}$$

Here the factor “3” represents the spatial dimensions of our investigated system.

The microcanonical partition function corresponding to the above non-Hamiltonian equation can be constructed using the metric $\exp(-w)$ and the energy conservation condition

$$\begin{aligned} Z = \int \prod_{i=0}^{N} \mathrm{d}\boldsymbol{p}_i \int \prod_{i=0}^{N} \mathrm{d}\boldsymbol{r}_i \int \mathrm{d}p_{N_e} \int \mathrm{d}N_e \int \mathrm{d}p_\eta \int \mathrm{d}\eta\, e^{(3N+1)\eta} \\ \times \delta\left(H_{\mathrm{sys}}(\{\boldsymbol{r}_i, \boldsymbol{p}_i\}, N_{\mathrm{e}}) + \frac{p_\eta^2}{2Q} + g k_{\mathrm{B}} T \eta + \frac{p_{N_{\mathrm{e}}}^2}{2m_{N_{\mathrm{e}}}} - \mu_{\mathrm{e}} N_{\mathrm{e}} - C_1\right) \end{aligned} \tag{S8}$$

where $H_{\mathrm{sys}}(\{\boldsymbol{r}_i, \boldsymbol{p}_i\}, N_{\mathrm{e}}) = \sum_{i=1}^{N} \frac{\boldsymbol{p}_i^2}{2m_i} + U(\{\boldsymbol{r}_i\}, N_{\mathrm{e}})$. To obtain the distribution function in the physical phase space, integration is performed over the variables $\eta$ and $p_\eta$. The integration over $\eta$ is carried out using the Dirac function, which imposes the condition that:

$$\eta = \frac{1}{gk_{\mathrm{B}}T}\left(C_1 - H_{\mathrm{sys}}(\{\boldsymbol{r}_i, \boldsymbol{p}_i\}, N_{\mathrm{e}}) - \frac{p_\eta^2}{2Q} - \frac{p_{N_{\mathrm{e}}}^2}{2m_{N_{\mathrm{e}}}} + \mu_{\mathrm{e}} N_{\mathrm{e}}\right) \tag{S9}$$

where $g = 3N + 1$ represents the number of degrees of freedom and $N$ is the number of particles. Substituting this result into the equation (**Eq. S8**),

$$\begin{aligned} Z &= e^{\beta C_1} \int \mathrm{d}p_\eta\, e^{-\beta \frac{p_\eta^2}{2Q}} \int \mathrm{d}p_{N_{\mathrm{e}}}\, e^{-\beta \frac{p_{N_{\mathrm{e}}}^2}{2m_{N_{\mathrm{e}}}}} \int \prod_{i=0}^{N} \mathrm{d}\boldsymbol{p}_i \int \prod_{i=0}^{N} \mathrm{d}\boldsymbol{r}_i \int \mathrm{d}N_{\mathrm{e}} e^{-\beta[H_{\mathrm{sys}}(\{\boldsymbol{r}_i, \boldsymbol{p}_i\}, N_{\mathrm{e}}) - \mu_{\mathrm{e}} N_{\mathrm{e}}]} \\ &= A \int \prod_{i=0}^{N} \mathrm{d}\boldsymbol{p}_i \int \prod_{i=0}^{N} \mathrm{d}\boldsymbol{r}_i \int \mathrm{d}N_{\mathrm{e}} e^{-\beta \Omega} = A\Xi(\beta, \mu_{\mathrm{e}}) \end{aligned} \tag{S10}$$

$$\Omega = \sum_{i=1}^{N} \frac{\boldsymbol{p}_i^2}{2m_i} + U(\{\boldsymbol{r}_i\}, N_{\mathrm{e}}) - \mu_{\mathrm{e}} N_{\mathrm{e}} \tag{S11}$$

The above derivation demonstrates that, aside from a constant prefactor $A$, the Nosé–Hoover type equations generate a grand canonical distribution in the physical subsystem variables $(\{\boldsymbol{p}_i\}, \{\boldsymbol{r}_i\}, N_{\mathrm{e}})$. Here the GC ensemble partition function $\Xi(\beta, \mu_{\mathrm{e}})$ follows the standard definition provided in the main text **Eq. 1**.

## S2. Discussion about beads' electron number $N_{\mathrm{e}}$ in the GC path integral formalism

Under the adiabatic approximation, the system's quantum state can be expressed as the direct product of $|n(\boldsymbol{R})\rangle|\boldsymbol{R}\rangle$, where $|\boldsymbol{R}\rangle$ represents the nuclear position state (corresponding to all particles' spatial coordinates) and $|n(\boldsymbol{R})\rangle$ denotes the electronic adiabatic state at the nuclear position $\boldsymbol{R}$. Considering only the ground state for electrons, the adiabatic state with a variable number of electrons simplifies to $|N_{\mathrm{e}}(\boldsymbol{R})\rangle$. With this simplification, the quantum partition function in the GC ensemble can be derived by adding a trace over the electronic adiabatic states.

$$\begin{aligned} \Xi_{\mathrm{qtm}}(\beta, \mu_{\mathrm{e}}) &= \mathrm{Tr}_{\mathrm{n}}\{\mathrm{Tr}_{\mathrm{e}}[\exp(-\beta(\widehat{\mathcal{H}} - \mu_{\mathrm{e}}\widehat{N}_{\mathrm{e}}))]\} \\ &= \lim_{P\to\infty} \mathrm{Tr}_{\mathrm{n}}\left\{\mathrm{Tr}_{\mathrm{e}}\left[\exp(-\beta(\widehat{\mathcal{H}} - \mu_{\mathrm{e}}\widehat{N}_{\mathrm{e}})/P)^P\right]\right\} \\ &= \lim_{P\to\infty} \int \mathrm{d}\boldsymbol{R}^{(1)} \dots \mathrm{d}\boldsymbol{R}^{(P)} \sum_{N_{\mathrm{e}}^{(1)} \dots N_{\mathrm{e}}^{(P)}} \langle \boldsymbol{R}^{(1)}| \left\langle N_{\mathrm{e}}^{(1)}(\boldsymbol{R}^{(1)})\right| \exp(-\beta(\widehat{\mathcal{H}} - \mu_{\mathrm{e}}\widehat{N}_{\mathrm{e}})/P) \left|N_{\mathrm{e}}^{(2)}(\boldsymbol{R}^{(2)})\right\rangle |\boldsymbol{R}^{(2)}\rangle \end{aligned} \tag{S12}$$

$$\times \cdots \times \langle \boldsymbol{R}^{(P-1)} | \left\langle N_{\mathrm{e}}^{(P-1)}(\boldsymbol{R}^{(P-1)}) \middle| \exp\left(-\beta(\widehat{\mathcal{H}} - \mu_{\mathrm{e}}\widehat{N}_{\mathrm{e}})/P\right) \middle| N_{\mathrm{e}}^{(P)}(\boldsymbol{R}^{(P)}) \right\rangle | \boldsymbol{R}^{(P)} \rangle$$

$$\times \langle \boldsymbol{R}^{(P)} | \left\langle N_{\mathrm{e}}^{(P)}(\boldsymbol{R}^{(P)}) \middle| \exp\left(-\beta(\widehat{\mathcal{H}} - \mu_{\mathrm{e}}\widehat{N}_{\mathrm{e}})/P\right) \middle| N_{\mathrm{e}}^{(1)}(\boldsymbol{R}^{(1)}) \right\rangle | \boldsymbol{R}^{(1)} \rangle$$

$$= \lim_{P\to\infty} \int \mathrm{d}\boldsymbol{R}^{(1)} \dots \mathrm{d}\boldsymbol{R}^{(P)} \sum_{N_{\mathrm{e}}^{(1)} \dots N_{\mathrm{e}}^{(P)}} \langle \boldsymbol{R}^{(1)} | \exp\left(-\beta \widehat{\mathcal{H}}_{N_{\mathrm{e}}^{(2)}}/P\right) | \boldsymbol{R}^{(2)} \rangle$$

$$\times \cdots \times \langle \boldsymbol{R}^{(P-1)} | \exp\left(-\beta \widehat{\mathcal{H}}_{N_{\mathrm{e}}^{(P)}}/P\right) | \boldsymbol{R}^{(P)} \rangle \langle \boldsymbol{R}^{(P)} | \exp\left(-\beta \widehat{\mathcal{H}}_{N_{\mathrm{e}}^{(1)}}/P\right) | \boldsymbol{R}^{(1)} \rangle$$

$$\times \exp\left(\beta\mu_{\mathrm{e}}/P \sum_{k=1}^{P} N_{\mathrm{e}}^{(k)}\right) \left\langle N_{\mathrm{e}}^{(1)}(\boldsymbol{R}^{(1)}) \middle| N_{\mathrm{e}}^{(2)}(\boldsymbol{R}^{(2)}) \right\rangle$$

$$\times \cdots \times \left\langle N_{\mathrm{e}}^{(P-1)}(\boldsymbol{R}^{(P-1)}) \middle| N_{\mathrm{e}}^{(P)}(\boldsymbol{R}^{(P)}) \right\rangle \left\langle N_{\mathrm{e}}^{(P)}(\boldsymbol{R}^{(P)}) \middle| N_{\mathrm{e}}^{(1)}(\boldsymbol{R}^{(1)}) \right\rangle$$

The electronic ground states corresponding to different electron numbers are orthogonal in the Fock space [1]. Consequently, as $P \to \infty$, the inner product satisfies the following equation:

$$\left\langle N_{\mathrm{e}}^{(i)}(\boldsymbol{R}^{(i)}) \middle| N_{\mathrm{e}}^{(j)}(\boldsymbol{R}^{(j)}) \right\rangle = \delta_{N_{\mathrm{e}}^{(i)}, N_{\mathrm{e}}^{(j)}} \tag{S13}$$

Building on this property, the number of electrons across all beads in the path integral (PI) formalism remains identical. As a result, the GC ensemble quantum partition function can be expressed in terms of weighted summation of canonical ensemble quantum partition functions.

$$\begin{aligned} \Xi_{\mathrm{qtm}}(\beta, \mu_{\mathrm{e}}) &= \sum_{N_{\mathrm{e}}} \exp(\beta\mu_{\mathrm{e}}N_{\mathrm{e}}) \lim_{P\to\infty} \int \mathrm{d}\boldsymbol{R}^{(1)} \dots \mathrm{d}\boldsymbol{R}^{(P)} \langle \boldsymbol{R}^{(1)} | \exp\left(-\beta \widehat{\mathcal{H}}_{N_{\mathrm{e}}}/P\right) | \boldsymbol{R}^{(2)} \rangle \times \cdots \\ &\quad \times \langle \boldsymbol{R}^{(P-1)} | \exp\left(-\beta \widehat{\mathcal{H}}_{N_{\mathrm{e}}}/P\right) | \boldsymbol{R}^{(P)} \rangle \langle \boldsymbol{R}^{(P)} | \exp\left(-\beta \widehat{\mathcal{H}}_{N_{\mathrm{e}}}/P\right) | \boldsymbol{R}^{(1)} \rangle \\ &= \sum_{N_{\mathrm{e}}} \exp(\beta\mu_{\mathrm{e}}N_{\mathrm{e}})\, Q_{\mathrm{qtm}}(\beta, N_{\mathrm{e}}) \end{aligned} \tag{S14}$$

## S3. Derivations of the TP-PIMD Method

Building upon the framework established by the TP-Classical MD, we first derive the thermostatted-potentiostatted path integral MD (TP-PIMD), which thermodynamically satisfies the quantum GC partition function. To construct the extended Hamiltonian of the TP-PIMD approach, we refer to the Nosé–Hoover thermostatted-PIMD method [2] and the TP-Classical MD algorithm [3]. We note that a simplified version of mathematical expressions for a single particle system is shown in this paper to ensure a clear presentation of dynamic equations' key ideas, while our code and simulations indeed deal with a three-dimensional multi-particle system. In the spatial representation, the grand partition function can be written as:

$$\Xi_{\mathrm{qtm}}(\beta,\mu_{\mathrm{e}}) = \int\prod_{j=1}^{P}\mathrm{d}\boldsymbol{p}^{(j)}\int\prod_{j=1}^{P}\mathrm{d}\boldsymbol{r}^{(j)}\int\mathrm{d}N_{\mathrm{e}}\,e^{-\beta\left[H_{\mathrm{qm}}\left(\{\boldsymbol{r}^{(j)},\boldsymbol{p}^{(j)}\},N_{\mathrm{e}}\right)-\mu_{\mathrm{e}}N_{\mathrm{e}}\right]} \tag{S15}$$

$$H_{\mathrm{qm}}\left(\{\boldsymbol{r}^{(j)},\boldsymbol{p}^{(j)}\},N_{\mathrm{e}}\right) = \sum_{j=1}^{P}\left[\frac{\left[\boldsymbol{p}^{(j)}\right]^2}{2m}+\frac{1}{2}m\omega_P^2\left[\boldsymbol{r}^{(j)}-\boldsymbol{r}^{(j-1)}\right]^2+\frac{1}{P}V(\boldsymbol{r}^{(j)},N_{\mathrm{e}})\right] \tag{S16}$$

with $\omega_P = \sqrt{P}/\beta\hbar$ and $\boldsymbol{r}^{(0)} \equiv \boldsymbol{r}^{(P)}$. $V(\boldsymbol{r}^{(j)},N_{\mathrm{e}})$ represents the potential energy of each bead. We define the effective potential energy function as:

$$U\left(\{\boldsymbol{r}^{(j)}\},N_{\mathrm{e}}\right) = \sum_{j=1}^{P}\frac{1}{P}V(\boldsymbol{r}^{(j)},N_{\mathrm{e}}) \tag{S17}$$

Employing the commonly used strategy to enhance the sampling efficiency of TP-PIMD, we perform a variable transformation from $\{\boldsymbol{r}^{(j)},\boldsymbol{p}^{(j)}\}$ $(j = 1,2,\dots,P)$ to normal mode variables $\{\tilde{\boldsymbol{r}}^{(k)},\tilde{\boldsymbol{p}}^{(k)}\}$ $(k = 0,1,\dots,P-1)$ [2,4,5]. By applying the normal mode transformation, the coupled harmonic terms are decoupled, simplifying the system to independent normal modes with distinct frequencies. This simplification enables stable integration and adequate sampling of all modes. The dynamic equations are then transformed from the beads representation to the normal modes representation [2,4].

$$\begin{aligned}\tilde{\boldsymbol{p}}^{(k)} &= \sum_{j=1}^{P}\boldsymbol{p}^{(j)}C^{jk}\\ \tilde{\boldsymbol{r}}^{(k)} &= \sum_{j=1}^{P}\boldsymbol{r}^{(j)}C^{jk}\end{aligned} \tag{S18}$$

where in the case of even $P$, the elements of the orthogonal transformation matrix $C$ are

$$C^{jk} = \begin{cases}\sqrt{\frac{1}{P}}, & k=0\\ \sqrt{\frac{2}{P}}\cos(2\pi jk/P), & 1\le k\le\frac{P}{2}-1\\ \sqrt{\frac{1}{P}}(-1)^j, & k=P/2\\ \sqrt{\frac{2}{P}}\sin(2\pi jk/P), & \frac{P}{2}+1\le k\le P-1\end{cases} \tag{S19}$$

Under the normal mode transformation, the effective potential energy changes its form from $U(\{\boldsymbol{r}^{(j)}\}, N_\mathrm{e})$ $(j = 1, 2, \dots, P)$ to $\widetilde{U}(\{\tilde{\boldsymbol{r}}^{(k)}\}, N_\mathrm{e})$ $(k = 0, 1, \dots, P-1)$, which is the expression referred as $\widetilde{U}$. All quantities with a tilde are those derived after performing the normal mode transformation and the extended quantum Hamiltonian becomes

$$\widetilde{H}_\mathrm{qm}(\{\tilde{\boldsymbol{r}}^{(k)}, \tilde{\boldsymbol{p}}^{(k)}\}, N_\mathrm{e}) = \sum_{k=0}^{P-1} \left[ \frac{[\tilde{\boldsymbol{p}}^{(k)}]^2}{2m} + \frac{1}{2} m [\tilde{\omega}^{(k)} \tilde{\boldsymbol{r}}^{(k)}]^2 \right] + \widetilde{U}(\{\tilde{\boldsymbol{r}}^{(k)}\}, N_\mathrm{e}) \tag{S20}$$

$$\widetilde{H}_\mathrm{qm}^\mathrm{ext} = \widetilde{H}_\mathrm{qm}(\{\tilde{\boldsymbol{r}}^{(k)}, \tilde{\boldsymbol{p}}^{(k)}\}, N_\mathrm{e}) + \sum_{k=0}^{P-1} \left[ \frac{p_{\eta^{(k)}}^2}{2Q^{(k)}} + g k_\mathrm{B} T \eta^{(k)} \right] + \frac{p_{N_\mathrm{e}}^2}{2m_{N_\mathrm{e}}} - \mu_\mathrm{e} N_\mathrm{e} \tag{S21}$$

with $\tilde{\omega}^{(k)} = 2\omega_P \sin(k\pi/P)$. We thus obtain the TP-PIMD equations of motion for the single-particle three-dimensional system described by (with $k = 0, 1, \dots, P-1$):

$$\begin{cases} \dot{\tilde{\boldsymbol{r}}}^{(k)} = \dfrac{\tilde{\boldsymbol{p}}^{(k)}}{m} \\ \dot{\tilde{\boldsymbol{p}}}^{(k)} = -m[\tilde{\omega}^{(k)}]^2 \tilde{\boldsymbol{r}}^{(k)} - \dfrac{\partial \widetilde{U}}{\partial \tilde{\boldsymbol{r}}^{(k)}} - \dfrac{p_{\eta^{(k)}}}{Q^{(k)}} \tilde{\boldsymbol{p}}^{(k)} \\ \dot{\eta}^{(k)} = \dfrac{p_{\eta^{(k)}}}{Q^{(k)}} \\ \dot{p}_{\eta^{(k)}} = \dfrac{[\tilde{\boldsymbol{p}}^{(k)}]^2}{m} - g k_\mathrm{B} T + \dfrac{1}{P} \dfrac{p_{N_\mathrm{e}}^2}{m_{N_\mathrm{e}}} \\ \dot{N}_\mathrm{e} = \dfrac{p_{N_\mathrm{e}}}{m_{N_\mathrm{e}}} \\ \dot{p}_{N_\mathrm{e}} = -\dfrac{\partial \widetilde{U}}{\partial N_\mathrm{e}} + \mu_\mathrm{e} - \dfrac{p_{N_\mathrm{e}}}{P} \displaystyle\sum_{k=0}^{P-1} \dfrac{p_{\eta^{(k)}}}{Q^{(k)}} \end{cases} \tag{S22}$$

We begin by identifying the conservation laws associated with the above dynamic equations.

$$
\begin{aligned}
\frac{d\widetilde{H}_{\mathrm{qm}}^{\mathrm{ext}}}{dt} &= \sum_{k=0}^{P-1}\left[\frac{\widetilde{\boldsymbol{p}}^{(k)}}{m}\dot{\widetilde{\boldsymbol{p}}}^{(k)} + m\left[\widetilde{\omega}^{(k)}\right]^2\widetilde{\boldsymbol{r}}^{(k)}\dot{\widetilde{\boldsymbol{r}}}^{(k)} + \frac{\partial\widetilde{U}}{\partial\widetilde{\boldsymbol{r}}^{(k)}}\dot{\widetilde{\boldsymbol{r}}}^{(k)}\right] + \frac{\partial\widetilde{U}}{\partial N_{\mathrm{e}}}\dot{N}_{\mathrm{e}} \\
&+ \sum_{k=0}^{P-1}\left[\frac{p_{\eta^{(k)}}}{Q^{(k)}}\dot{p}_{\eta^{(k)}} + gk_{\mathrm{B}}T\dot{\eta}^{(k)}\right] + \frac{p_{N_{\mathrm{e}}}}{m_{N_{\mathrm{e}}}}\dot{p}_{N_{\mathrm{e}}} - \mu_{\mathrm{e}}\dot{N}_{\mathrm{e}} \\
&= \sum_{k=0}^{P-1}\left[\frac{\widetilde{\boldsymbol{p}}^{(k)}}{m}\left(-m\left[\widetilde{\omega}^{(k)}\right]^2\widetilde{\boldsymbol{r}}^{(k)} - \frac{\partial\widetilde{U}}{\partial\widetilde{\boldsymbol{r}}^{(k)}} - \frac{p_{\eta^{(k)}}}{Q^{(k)}}\widetilde{\boldsymbol{p}}^{(k)}\right) + m\left[\widetilde{\omega}^{(k)}\right]^2\widetilde{\boldsymbol{r}}^{(k)}\frac{\widetilde{\boldsymbol{p}}^{(k)}}{m}\right. \\
&\left. + \frac{\partial\widetilde{U}}{\partial\widetilde{\boldsymbol{r}}^{(k)}}\frac{\widetilde{\boldsymbol{p}}^{(k)}}{m}\right] + \frac{\partial\widetilde{U}}{\partial N_{\mathrm{e}}}\frac{p_{N_{\mathrm{e}}}}{m_{N_{\mathrm{e}}}} \\
&+ \sum_{k=0}^{P-1}\left[\frac{p_{\eta^{(k)}}}{Q^{(k)}}\left(\frac{[\widetilde{\boldsymbol{p}}^{(k)}]^2}{m} - gk_{\mathrm{B}}T + \frac{1}{P}\frac{p_{N_{\mathrm{e}}}^2}{m_{N_{\mathrm{e}}}}\right) + gk_BT\frac{p_{\eta^{(k)}}}{Q^{(k)}}\right] \\
&+ \frac{p_{N_{\mathrm{e}}}}{m_{N_{\mathrm{e}}}}\left(-\frac{\partial\widetilde{U}}{\partial N_{\mathrm{e}}} + \mu_{\mathrm{e}} - \frac{p_{N_{\mathrm{e}}}}{P}\sum_{k=0}^{P-1}\frac{p_{\eta^{(k)}}}{Q^{(k)}}\right) - \mu_{\mathrm{e}}\frac{p_{N_{\mathrm{e}}}}{m_{N_{\mathrm{e}}}} = 0
\end{aligned}
\tag{S23}
$$

We then verify that sampled microstates' distribution conforms to the grand canonical ensemble. We first need to compute the phase-space compressibility as follows:

$$
\begin{aligned}
\kappa &= \sum_{k=0}^{P-1}\left[\frac{\partial\dot{\widetilde{\boldsymbol{r}}}^{(k)}}{\partial\widetilde{\boldsymbol{r}}^{(k)}} + \frac{\partial\dot{\widetilde{\boldsymbol{p}}}^{(k)}}{\partial\widetilde{\boldsymbol{p}}^{(k)}} + \frac{\partial\dot{\eta}^{(k)}}{\partial\eta^{(k)}} + \frac{\partial\dot{p}_{\eta^{(k)}}}{\partial p_{\eta^{(k)}}}\right] + \frac{\partial\dot{N}_{\mathrm{e}}}{\partial N_{\mathrm{e}}} + \frac{\partial\dot{p}_{N_{\mathrm{e}}}}{p_{N_{\mathrm{e}}}} = \sum_{k=0}^{P-1}\left(-\frac{p_{\eta^{(k)}}}{Q^{(k)}}\right) - \frac{1}{P}\sum_{k=0}^{P-1}\frac{p_{\eta^{(k)}}}{Q^{(k)}} \\
&= -\left(1+\frac{1}{P}\right)\sum_{k=0}^{P-1}\frac{p_{\eta^{(k)}}}{Q^{(k)}} = -\left(1+\frac{1}{P}\right)\sum_{k=0}^{P-1}\dot{\eta}^{(k)}
\end{aligned}
\tag{S24}
$$

Here, $\exp(-w)$ is the extended phase-space metric, and the relation between $w$ and $\kappa$ is given by **Eq. S6**. Specifically, we have:

$$
\exp(-w) = \exp\,[(1+\frac{1}{P})\sum_{k=0}^{P-1}\eta^{(k)}]
\tag{S25}
$$

The microcanonical partition function can be constructed using the metric $\exp(-w)$ and the energy conservation condition:

$$Z = \int \prod_{k=0}^{P-1} \mathrm{d}\widetilde{\boldsymbol{p}}^{(k)} \int \prod_{k=0}^{P-1} \mathrm{d}\tilde{\boldsymbol{r}}^{(k)} \int \prod_{k=0}^{P-1} \mathrm{d}p_{\eta^{(k)}} \int \prod_{k=0}^{P-1} \mathrm{d}\eta^{(k)} \int dp_{N_\mathrm{e}} \int dN_\mathrm{e} \exp\left[\left(1 + \frac{1}{P}\right) \sum_{k=0}^{P-1} \eta^{(k)}\right] \times \delta \left\{ \sum_{k=0}^{P-1} \left[ \frac{\left[\widetilde{\boldsymbol{p}}^{(k)}\right]^2}{2m} + \frac{1}{2} m \left[\widetilde{\omega}^{(k)} \tilde{\boldsymbol{r}}^{(k)}\right]^2 + \frac{p_{\eta^{(k)}}^2}{2Q^{(k)}} + g k_\mathrm{B} T \eta^{(k)} \right] + \widetilde{U}\left(\{\tilde{\boldsymbol{r}}^{(k)}\}, N_\mathrm{e}\right) + \frac{p_{N_\mathrm{e}}^2}{2m_{N_\mathrm{e}}} - \mu_\mathrm{e} N_\mathrm{e} - C_1 \right\} \quad \text{(S26)}$$

where the term $\eta^{(P-1)}$ is given by:

$$\eta^{(P-1)} = -\sum_{k=0}^{P-2} \eta^{(k)} + \frac{1}{g k_\mathrm{B} T} \left\{ -\sum_{k=0}^{P-1} \left[ \frac{\left[\widetilde{\boldsymbol{p}}^{(k)}\right]^2}{2m} + \frac{1}{2} m \left[\widetilde{\omega}^{(k)} \tilde{\boldsymbol{r}}^{(k)}\right]^2 + \frac{p_{\eta^{(k)}}^2}{2Q^{(k)}} \right] - \widetilde{U}\left(\{\tilde{\boldsymbol{r}}^{(k)}\}, N_\mathrm{e}\right) - \frac{p_{N_\mathrm{e}}^2}{2m_{N_\mathrm{e}}} + \mu_\mathrm{e} N_\mathrm{e} + C_1 \right\} \quad \text{(S27)}$$

Substituting this expression and $g = 1 + \frac{1}{P}$ into the equation (**Eq. S25**), we obtain:

$$\begin{aligned} \exp(-w) &= \exp\left[\left(1 + \frac{1}{P}\right)\left(\sum_{k=0}^{P-2} \eta^{(k)} + \eta^{(P-1)}\right)\right] \\ &= \exp\left[ -\beta \left\{ \sum_{k=0}^{P-1} \left[ \frac{\left[\widetilde{\boldsymbol{p}}^{(k)}\right]^2}{2m} + \frac{1}{2} m \left[\widetilde{\omega}^{(k)} \tilde{\boldsymbol{r}}^{(k)}\right]^2 + \frac{p_{\eta^{(k)}}^2}{2Q^{(k)}} \right] + \widetilde{U}\left(\{\tilde{\boldsymbol{r}}^{(k)}\}, N_\mathrm{e}\right) + \frac{p_{N_\mathrm{e}}^2}{2m_{N_\mathrm{e}}} - \mu_\mathrm{e} N_\mathrm{e} - C_1 \right\} \right] \end{aligned} \quad \text{(S28)}$$

Substituting this result into the equation (**Eq. S26**), we then have:

$$
\begin{aligned}
Z &= \mathrm{e}^{\beta C_1} \int \prod_{k=0}^{P-2} \mathrm{d}\eta^{(k)} \int \prod_{k=0}^{P-1} \mathrm{d}p_{\eta^{(k)}}\, e^{-\beta \frac{p_{\eta^{(k)}}^2}{2Q^{(k)}}} \int \mathrm{d}p_{N_\mathrm{e}} e^{-\beta \frac{p_{N_\mathrm{e}}^2}{2m_{N_\mathrm{e}}}} \\
&\times \int \mathrm{d}N_\mathrm{e} \int \prod_{k=0}^{P-1} \mathrm{d}\tilde{\boldsymbol{p}}^{(k)} \int \prod_{k=0}^{P-1} \mathrm{d}\tilde{\boldsymbol{r}}^{(k)} \exp\left[-\beta \left\{ \sum_{k=0}^{P-1} \left[ \frac{\left[\tilde{\boldsymbol{p}}^{(k)}\right]^2}{2m} + \frac{1}{2} m \left[\tilde{\omega}^{(k)} \tilde{\boldsymbol{r}}^{(k)}\right]^2 \right] + \tilde{U}\left(\{\tilde{\boldsymbol{r}}^{(k)}\}, N_\mathrm{e}\right) - \mu_\mathrm{e} N_\mathrm{e} \right\}\right]
\end{aligned}
\tag{S29}
$$

$$
\begin{aligned}
&= \mathrm{e}^{\beta C_1} \int \prod_{k=0}^{P-2} \mathrm{d}\eta^{(k)} \int \prod_{k=0}^{P-1} \mathrm{d}p_{\eta^{(k)}}\, e^{-\beta \frac{p_{\eta^{(k)}}^2}{2Q^{(k)}}} \int \mathrm{d}p_{N_\mathrm{e}} e^{-\beta \frac{p_{N_\mathrm{e}}^2}{2m_{N_\mathrm{e}}}} \int \prod_{k=0}^{P-1} \mathrm{d}\tilde{\boldsymbol{p}}^{(k)}\, e^{-\beta \sum_{k=0}^{P-1} \frac{\left[\tilde{\boldsymbol{p}}^{(k)}\right]^2}{2m}} \\
&\times \int \mathrm{d}N_\mathrm{e}\, e^{\beta \mu_\mathrm{e} N_\mathrm{e}} \int \prod_{k=0}^{P-1} \mathrm{d}\tilde{\boldsymbol{r}}^{(k)} \exp\left[-\beta \left\{ \sum_{k=0}^{P-1} \frac{1}{2} m \left[\tilde{\omega}^{(k)} \tilde{\boldsymbol{r}}^{(k)}\right]^2 + \tilde{U}\left(\{\tilde{\boldsymbol{r}}^{(k)}\}, N_\mathrm{e}\right) \right\}\right] = A \Xi_{\mathrm{qtm}}(\beta, \mu_\mathrm{e})
\end{aligned}
$$

$$
\Xi_{\mathrm{qtm}}(\beta, \mu_\mathrm{e}) = \int \mathrm{d}N_\mathrm{e}\, e^{\beta \mu_\mathrm{e} N_\mathrm{e}} \int \prod_{k=0}^{P-1} \mathrm{d}\tilde{\boldsymbol{r}}^{(k)} \exp\left[-\beta \left\{ \sum_{k=0}^{P-1} \frac{1}{2} m \left[\tilde{\omega}^{(k)} \tilde{\boldsymbol{r}}^{(k)}\right]^2 + \tilde{U}\left(\{\tilde{\boldsymbol{r}}^{(k)}\}, N_\mathrm{e}\right) \right\}\right] \tag{S30}
$$

The above derivation demonstrates that, aside from a constant prefactor $A$, the Nosé–Hoover type equations generate a grand canonical distribution in the physical subsystem variables $\left(\{\tilde{\boldsymbol{r}}^{(k)}\}, N_\mathrm{e}\right)$.

## S4. Derivations of the thermostatted-potentiostatted ring-polymer MD equations

PIMD satisfies the thermodynamic distribution and is suitable for configurational sampling, but cannot be directly used for studying dynamics. [2] Ring polymer molecular dynamics (RPMD) [6] is employed to compute quantum dynamic correlation functions. To better account for the thermostat, previous studies developed T-RPMD, which is kind of a combination of centroid molecular dynamics (CMD) [7] and RPMD, ensuring sufficient phase space sampling [8,9]. To preserve dynamics, thermostats are applied only to the internal modes, excluding the first normal mode (known as the centroid mode). The corresponding modification for our dynamic equations is straightforward: we can simply remove the thermostat coupling from the centroid mode. If the system is quantized into $P$ beads in PI, the resulting dynamic equations of our proposed thermostated-potentiostated RPMD (TP-RPMD) are shown below (a representative single-particle system for clarity):

$$\begin{cases} \dot{\tilde{\boldsymbol{r}}}^{(k)} = \dfrac{\tilde{\boldsymbol{p}}^{(k)}}{m}, \qquad k = 0,1,\dots,P-1 \\ \dot{\tilde{\boldsymbol{p}}}^{(k)} = -m\left[\tilde{\omega}^{(k)}\right]^2 \tilde{\boldsymbol{r}}^{(k)} - \dfrac{\partial \tilde{U}}{\partial \tilde{\boldsymbol{r}}^{(k)}}, \qquad k = 0 \\ \dot{\tilde{\boldsymbol{p}}}^{(k)} = -m\left[\tilde{\omega}^{(k)}\right]^2 \tilde{\boldsymbol{r}}^{(k)} - \dfrac{\partial \tilde{U}}{\partial \tilde{\boldsymbol{r}}^{(k)}} - \dfrac{p_{\eta^{(k)}}}{Q^{(k)}} \tilde{\boldsymbol{p}}^{(k)}, \qquad k = 1,\dots,P-1 \\ \dot{\eta}^{(k)} = \dfrac{p_{\eta^{(k)}}}{Q^{(k)}}, \qquad k = 1,\dots,P-1 \\ \dot{p}_{\eta^{(k)}} = \dfrac{[\tilde{\boldsymbol{p}}^{(k)}]^2}{m} - g k_{\mathrm{B}} T + \dfrac{1}{P-1} \dfrac{p_{N_{\mathrm{e}}}^2}{m_{N_{\mathrm{e}}}}, \qquad k = 1,\dots,P-1 \\ \dot{N}_{\mathrm{e}} = \dfrac{p_{N_{\mathrm{e}}}}{m_{N_{\mathrm{e}}}} \\ \dot{p}_{N_{\mathrm{e}}} = -\dfrac{\partial \tilde{U}}{\partial N_{\mathrm{e}}} + \mu_{\mathrm{e}} - \dfrac{p_{N_{\mathrm{e}}}}{P-1} \displaystyle\sum_{k=1}^{P-1} \dfrac{p_{\eta^{(k)}}}{Q^{(k)}} \end{cases} \tag{S31}$$

where the superscript $k$ refers to the normal mode index. $\eta^{(k)}$ and $p_{\eta^{(k)}}$ represent the generalized coordinates and generalized momentums associated with the degrees of freedom of the Nosé–Hoover thermostats, which are attached to the $k$-th normal mode of the ring-polymer. Note that the thermostat is not applied to the centroid mode ($k = 0$), analogous to the strategy of the Centroid MD algorithm [7]. $N_{\mathrm{e}}$ is the system's total number of electrons, and $p_{N_{\mathrm{e}}}$ is the associated generalized momentum. Since all beads in the PI formulation share the same $N_{\mathrm{e}}$ value (refer to **SI Section 2** for details), the quantized system in TP-RPMD is coupled to a single potentiostat. $\tilde{\omega}^{(k)}$ characterizes the strength of the harmonic interaction between neighboring beads in the PI formulation.

We actually use the Nosé–Hoover chain strategy in our simulations to achieve a more efficient implementation of thermostats [2]. If we add $M$ additional thermostat variable pairs $\eta_\gamma^{(k)}$ and $p_{\eta_\gamma^{(k)}}$, where $\gamma = 1,\dots,M$, with the subscript $\gamma$ indicating the Nosé–Hoover chain index, the equations of motion can be expressed as:

$$
\begin{cases}
\dot{\tilde{\boldsymbol{r}}}^{(k)} = \dfrac{\tilde{\boldsymbol{p}}^{(k)}}{m}, \qquad k = 0,1,\dots,P-1 \\
\dot{\tilde{\boldsymbol{p}}}^{(k)} = -m\left[\tilde{\omega}^{(k)}\right]^2 \tilde{\boldsymbol{r}}^{(k)} - \dfrac{\partial \tilde{U}}{\partial \tilde{\boldsymbol{r}}^{(k)}}, \qquad k = 0 \\
\dot{\tilde{\boldsymbol{p}}}^{(k)} = -m\left[\tilde{\omega}^{(k)}\right]^2 \tilde{\boldsymbol{r}}^{(k)} - \dfrac{\partial \tilde{U}}{\partial \tilde{\boldsymbol{r}}^{(k)}} - \dfrac{p_{\eta_1^{(k)}}}{Q^{(k)}} \tilde{\boldsymbol{p}}^{(k)}, \qquad k = 1,\dots,P-1 \\
\dot{\eta}_\gamma^{(k)} = \dfrac{p_{\eta_\gamma^{(k)}}}{Q^{(k)}}, \qquad k = 1,\dots,P-1, \qquad \gamma = 1,\dots,M \\
\dot{p}_{\eta_1^{(k)}} = \dfrac{[\tilde{\boldsymbol{p}}^{(k)}]^2}{m} - g k_{\mathrm{B}} T + \dfrac{1}{P-1} \dfrac{p_{N_\mathrm{e}}^2}{m_{N_\mathrm{e}}} - \dfrac{p_{\eta_2^{(k)}}}{Q^{(k)}} p_{\eta_1^{(k)}}, \qquad k = 1,\dots,P-1 \\
\dot{p}_{\eta_\gamma^{(k)}} = \dfrac{p^2_{\eta_{\gamma-1}^{(k)}}}{Q^{(k)}} - g k_{\mathrm{B}} T - \dfrac{p_{\eta_{\gamma+1}^{(k)}}}{Q^{(k)}} p_{\eta_\gamma^{(k)}}, \qquad k = 1,\dots,P-1, \qquad \gamma = 2,\dots,M-1 \\
\dot{p}_{\eta_M^{(k)}} = \dfrac{p^2_{\eta_{M-1}^{(k)}}}{Q^{(k)}} - g k_{\mathrm{B}} T, \qquad k = 1,\dots,P-1 \\
\dot{N}_\mathrm{e} = \dfrac{p_{N_\mathrm{e}}}{m_{N_\mathrm{e}}} \\
\dot{p}_{N_\mathrm{e}} = -\dfrac{\partial \tilde{U}}{\partial N_\mathrm{e}} + \mu_\mathrm{e} - \dfrac{p_{N_\mathrm{e}}}{P-1} \displaystyle\sum_{k=1}^{P-1} \dfrac{p_{\eta_1^{(k)}}}{Q^{(k)}}
\end{cases}
\tag{S32}
$$

## S5. Setup of DFT calculations and MLFF training

Density functional theory (DFT) [10] calculations are used to label data for training the machine learning force fields (MLFF) model. We perform DFT single-point calculations using the first-principles package Atomic-orbital Based Ab-initio Computation at UStc (ABACUS) [11,12], which is well-suited for performing efficient DFT calculations on large-size supercells due to its feature of numerical atomic orbitals (NAO) basis set. We use norm-conserving pseudopotentials with valence configurations defined as [H]$1s^1$, [O]$2s^2 2p^4$, and [Pt] $5s^2 5p^6 6d^8 6s^2$. The Perdew-Burke-Ernzerhof (PBE) generalized gradient approximation (GGA) describes exchange-correlation interactions [13], while Grimme's D3 dispersion correction [14] accounts for van der Waals forces. We employ 2s1p, 2s2p1d, and 4s2p2d1f numerical atomic orbitals (NAOs) as basis sets, with cutoff radii of 6, 7, and 7 Bohr for H, O, and Pt, respectively. Periodic boundary condition (PBC) is applied to simulate the Pt/$H_2O$ interface using a 14.06 Å × 14.06 Å × 30.89 Å supercell. The Brillouin zone is sampled by a 2×2×1 k-point mesh, and a plane-wave kinetic energy cutoff of 100 Ry (1360 eV) is employed. We apply Gaussian smearing with a width of 0.02 Ry. A dipole correction [15] is included to account for net dipoles along the z direction of the supercell.

The MLFF is constructed using the DP-GEN [16] workflow, which involves three major steps: training the force field, exploring the extended ($\{\boldsymbol{r}_i\}$,$N_\mathrm{e}$) configurational space, and labeling new configurations via first-principles calculations. In our work, we adopt a modified version of the Deep Potential (DP) framework, referred to as DP-$N_\mathrm{e}$, which was developed in our previous study [17]. DP-$N_\mathrm{e}$ is an MLFF specifically adapted for electrochemical simulations, where the total electron number of the interfacial system is an additional degree of freedom in the force field to facilitate GC sampling (as can be seen by **Eq. 1**). The initial version of the MLFF is inherited from our previous work [17], where good performance has been demonstrated in the Pt/$H_2O$ interfacial system for HER study. In this work, we further supplement the training dataset by incorporating new configurations from classical/quantum dynamics trajectories (i.e., TP-classical MD and TP-RPMD simulations). DeepMD-kit is used for the training process [18–23].

The DP-GEN workflow iterates through training, exploration, and labeling. During the training process, four models, each initialized with different random seeds, are trained using DeePMD-kit for $4\times10^5$ steps. The embedding network consists of three layers with 25, 50, and 100 nodes, while the fitting network comprises three layers, each containing 240 nodes. The loss functions are optimized with an exponentially decaying learning rate, starting at $1.00\times10^{-3}$ and gradually decreasing to $3.51\times10^{-8}$. In the exploration process, configurations exhibiting model deviations between 0.10 and 0.60 eV/Å are selected as candidate structures. These candidates are subsequently labeled through *ab-initio* calculations and incorporated into the dataset for the next iteration. The DP-GEN iterations are considered converged when over 85% of the snapshots in a $10^4$-step trajectory exhibit model deviation below 0.10 eV/Å.

For testing the performance of our trained MLFF, 100 snapshots from a classical simulation trajectory (TP-Classical MD) and 100 snapshots from a quantum simulation trajectory (TP-RPMD) are selected to compare the energies and forces inferred by our DP-$N_\mathrm{e}$ MLFF [17] with those predicted by density functional theory (DFT) calculations (**Fig. S1**).

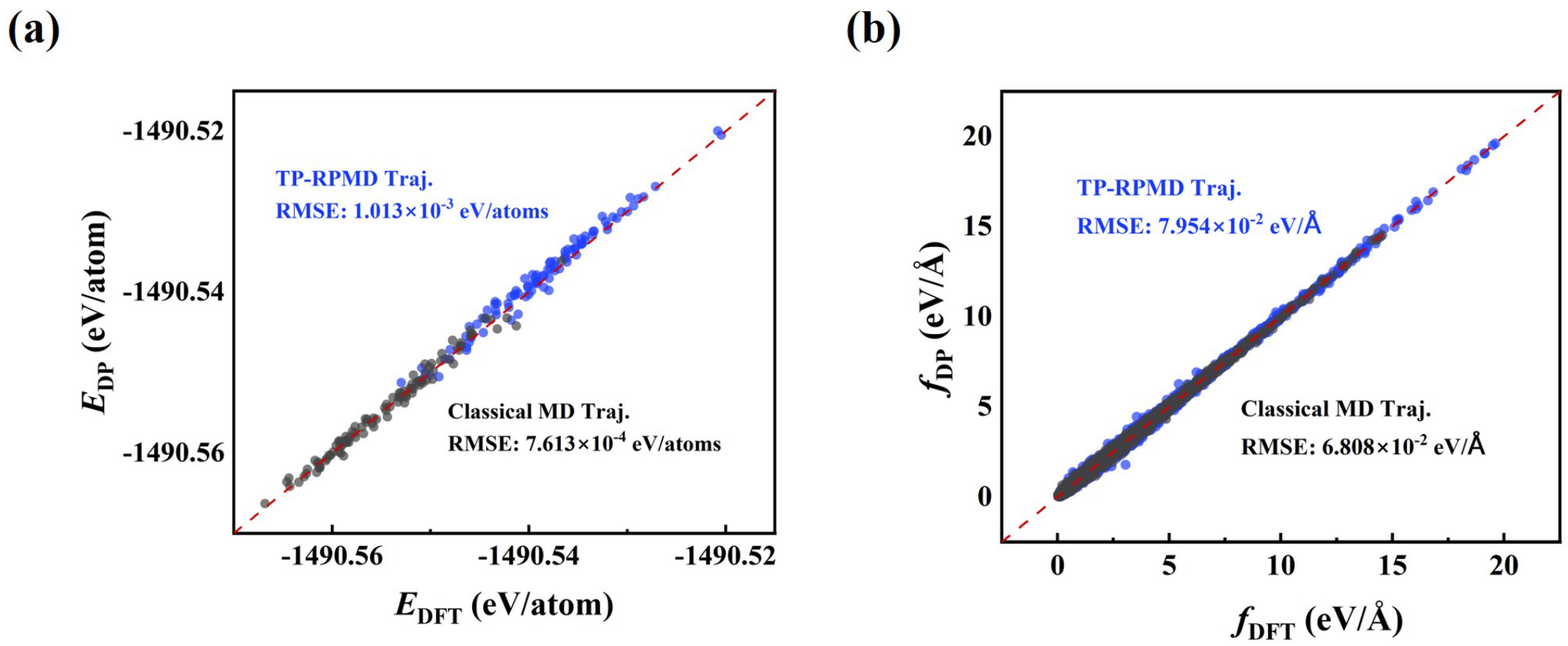

**Figure S1** Comparisons of energies ($E_{\mathrm{DP}}$ vs $E_{\mathrm{DFT}}$) and forces ($f_{\mathrm{DP}}$ vs $f_{\mathrm{DFT}}$) obtained by our DP-$N_{\mathrm{e}}$ and DFT calculations on the testing dataset. The total root mean square errors (RMSE) of energies and forces are listed in the panels.

## S6. Probability evaluation of the system being in final state along MD trajectories

The probability of our modeled system being in proton-coupled electron transfer (PCET) final state (FS) during the reaction process is assessed over 100 successive steps. The FS probability is calculated as the proportion of time the system spends in the FS region during the preceding 100 consecutive steps. The system is considered as truly residing in the FS region when the FS probability exceeds 50%, indicating that the PCET reaction is approximately "complete".

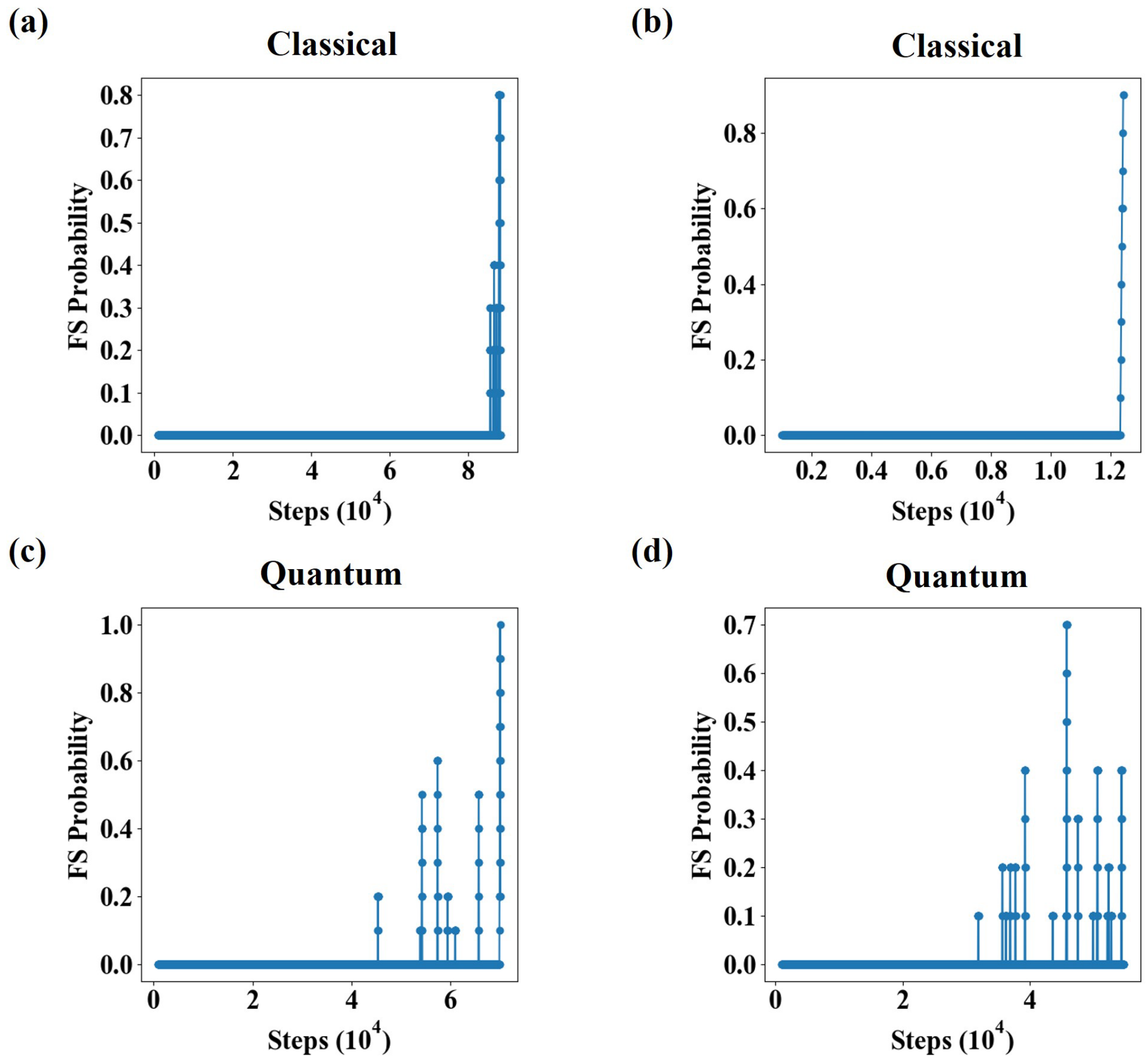

**Figure S2** (a)(b) Classical cases. (c)(d) Quantum cases. The four cases correspond to different trajectories with $\mu_e$ = -2.0 eV vs vacuum. The *x*-axis represents the MD time steps, and the *y*-axis shows the probability of the system staying in the FS region during the preceding 100 consecutive steps.

## S7. Selection of an optimal time range for linear fitting reaction rate constants

To select an appropriate time range for linear fitting reaction rate constants (**Fig. 3c-d** in the main text), we evaluate the fitting quality using the Pearson correlation coefficient [24]. In practice, we perform multiple fittings with different choices of $t^*$ and select the one resulting in the highest $R^2$ value.

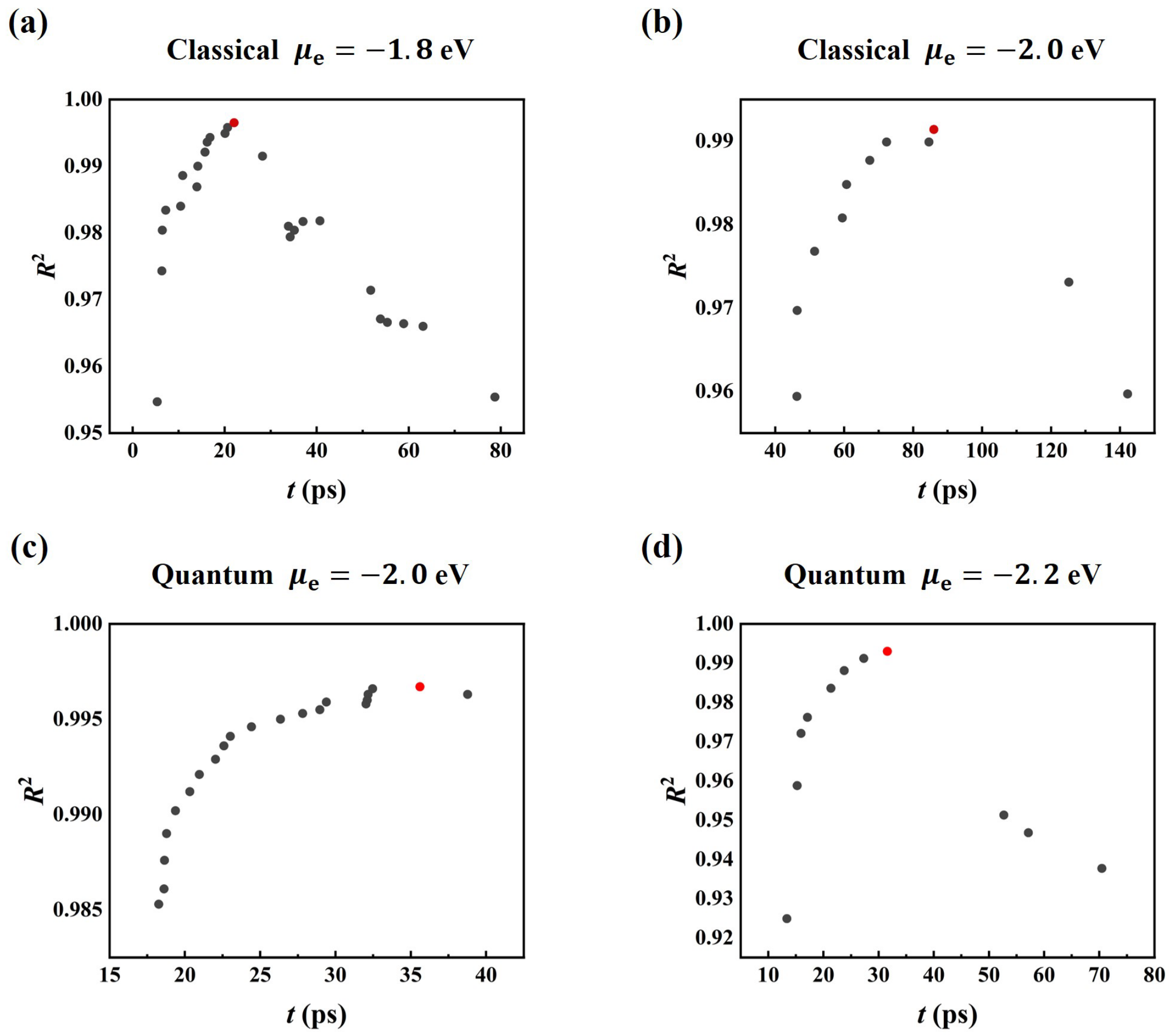

**Figure S3** Plots of scattering data points showing the relationship between $R^2$ and time $t$. The point where $R^2$ reaches its maximum is highlighted in red. The corresponding value of $t$ at this point is defined as $t^*$, which is chosen to provide the linear fitting of $\ln[1 - C_{AB}(t)]$ within the time range of $[0, t^*]$. (a)(b) represent the classical case, with $\mu_e$ = -1.8 eV and $\mu_e$ = -2.0 eV, and (c)(d) represent the quantum case, with $\mu_e$ = -2.0 eV and $\mu_e$ = -2.2 eV.

## S8. Validation of constant temperature and potential conditions in MD simulations

To check our implementation of constant temperature and constant potential conditions, we show the fluctuations and statistical averages of instantaneous temperature and work function (i.e. $-\partial U/\partial N_e$) values along TP-Classical MD and TP-RPMD trajectories, and compare with the macroscopic values of temperature $T$ and electrochemical potential $\mu_e$ set for the MD runs. For the quantum cases (**Fig. S4c-d**), the temperature in panel (c) corresponds to the average over internal normal modes (modes 1-15), as the thermostats are coupled to these modes, excluding the centroid mode. The work function values in panel (d) are calculated via beads average across all 16 replicas of the ring polymer. We can see in **Fig. S4** that the statistical averages of those microscopic quantities well converge to the preset macroscopic values, demonstrating successful achievement of constant temperature and constant potential conditions in our developed MD algorithm in this work.

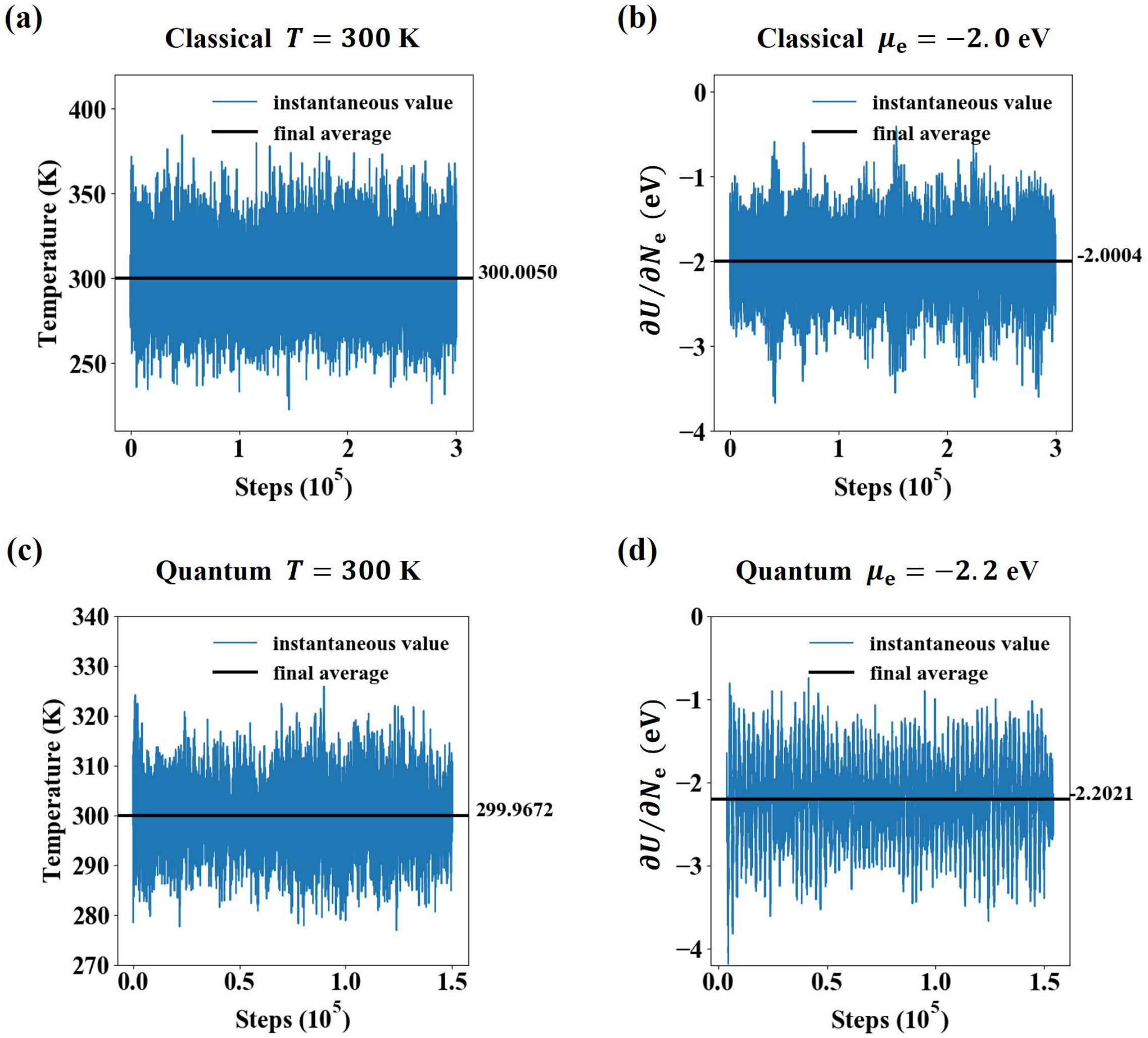

**Figure S4** (a) and (b) show the instantaneous values and final averages of temperature and $\partial U/\partial N_e$in classical situations (TP-Classical MD trajectories), with $T$ = 300 K and $\mu_e$ = -2.0 eV. (c) and (d) display the instantaneous values and final averages of modes-average temperature and beads-average $\partial U/\partial N_e$ in the quantum situations (TP-RPMD trajectories), with $T$ = 300 K and $\mu_e$ = -2.2 eV.

## Supplementary references